\documentclass[10pt]{article}

\usepackage[utf8]{inputenc}
\usepackage[T1]{fontenc}
\usepackage{lmodern}
\usepackage{amsmath, amssymb}
\usepackage{graphicx}
\usepackage{hyperref}
\usepackage{geometry}
\usepackage{authblk}  % For author and affiliation blocks
\usepackage{booktabs} 
\usepackage{tabularray}
\title{Negative Ties Highlight Hidden Extremes \\in Social Media Polarization}

\author[1,2,*]{E. Candellone}
\author[3,*]{S. A. Babul}
\author[1]{Ö. Togay}
\author[4,5]{A. Bovet}
\author[1,2,6]{J. Garcia-Bernardo}

\affil[1]{Department of Methodology and Statistics, Utrecht University, Utrecht, Netherlands}
\affil[2]{Centre for Complex Systems Studies, Utrecht University, Utrecht, Netherlands}
\affil[3]{Mathematical Institute, University of Oxford, United Kingdom}
\affil[4]{Department of Mathematical Modeling and Machine Learning, University of Zurich, Switzerland}
\affil[5]{Digital Society Initiative, University of Zurich, Switzerland}
\affil[6]{ODISSEI SoDa Team, Utrecht University, Utrecht, Netherlands}
\affil[*]{Joint first authors}

\begin{document}

\maketitle

\begin{abstract}
Human interactions in the online world comprise a combination of positive and negative exchanges. These diverse interactions can be captured using signed network representations, where edges take positive or negative weights to indicate the sentiment of the interaction between individuals. Signed networks offer valuable insights into online political polarization by capturing antagonistic interactions and ideological divides on social media platforms. This study analyzes polarization on Menéame, a Spanish social media platform that facilitates engagement with news stories through comments and voting.  Using a dual-method approach---Signed Hamiltonian Eigenvector Embedding for Proximity (SHEEP) for signed networks and Correspondence Analysis (CA) for unsigned networks---we investigate how including negative ties enhances the understanding of structural polarization levels across different conversation topics on the platform. While the unsigned Menéame network effectively delineates ideological communities, only by incorporating negative ties can we identify ideologically extreme users who engage in antagonistic behaviors: without them, the most extreme users remain indistinguishable from their less confrontational ideological peers.
\end{abstract}

\section{Introduction}\label{sec:introduction}
Online social networks have changed how people interact with news content, stay informed about current events, and form opinions on related topics (\cite{marchi2012facebook}).  This new mechanism for communication and information spread plays a key role in facilitating an increased polarization around controversial issues and amplifying political divisions and conflicts (\cite{garimella2017reducing, adamic2005political, conover2011political, hohmannQuantifyingIdeologicalPolarization2023a, barberaSocialMediaEcho2020, falkenberg_growing_2022,flamino_political_2023, tornberg2022digital}). As a consequence, political polarization is increasing not only within small, active partisan groups (\cite{nealSignTimesWeak2020a}) but also spreading more widely among the general population (\cite{nealSignTimesWeak2020a, abramowitzPolarizationMyth2008, abramowitzWhyCantWe2005}).\par

In political science literature, a distinction is made between ideological, affective, and structural (or interactional) polarization (\cite{hohmannQuantifyingIdeologicalPolarization2023a, dimaggioHaveAmericansSocial1996, bramsonUnderstandingPolarizationMeanings2017, lelkes2016mass, barberaTweetingLeftRight2015a, nealSignTimesWeak2020a, esauDestructivePolarizationDigital, adamic2005political,salloumSeparatingPolarizationNoise2021,yarchiPoliticalPolarizationDigital2020}). Ideological polarization refers to the widening gap between the political beliefs of different groups; affective polarization captures the emotional hostility and negative attitudes between political factions; and structural polarization examines the division of social interactions into homogeneous groups with minimal cross-group engagement. Given the recent increase in data available on online interactions between social media users, structural polarization (\cite{yarchiPoliticalPolarizationDigital2020, salloumSeparatingPolarizationNoise2021,falkenbergAffectiveInteractionalPolarization2023}) has become a rich area of study over the past few decades. Prior research has shown that social media platforms may amplify polarization by promoting the alignment of differences, facilitating the formation of echo chambers (clusters of like-minded users) and the rapid diffusion of biased information (\cite{delvicarioSpreadingMisinformationOnline2016,cinelliEchoChamberEffect2021a,barberaSocialMediaEcho2020, ferrazdearrudaModellingHowSocial2022a}). Ideological polarization can also be recovered from the network of interactions between social media users, under the assumption that users are more likely to interact with other users (or content) with whom they share a similar ideological stance (\cite{barberaTweetingLeftRight2015a, ribeiro2017everything}). The ideal points model, used to measure both voting behaviors (\cite{enelowSpatialTheoryVoting1984, pooleSpatialModelLegislative1985,pooleSpatialModelsParliamentary2005, clintonStatisticalAnalysisRoll2004, yuSpatialVotingModels2021}) and ideology (\cite{moodyPortraitPoliticalParty2013, waughPartyPolarizationCongress2009, barberaBirdsSameFeather2015}), is based on the same assumption: the probability of voting (positively or negatively) on some content depends on the latent ideological difference between the individual and the content. The model has been applied across various contexts, from a legislator voting on a piece of legislation to a user voting on a social media post. \par

Despite extensive research, studies of online polarization remain constrained by limited data. Social media platforms primarily provide information on positive interactions (e.g., likes or retweets on Twitter/X) or neutral interactions (e.g., mentions on Twitter/X), while negative interactions (e.g., downvotes on Reddit) are either unrecorded or only accessible to researchers at an aggregated level. As a result, most prior studies analyze structural polarization by constructing unsigned networks from online interactions, failing to distinguish between interactions that are positive or negative in nature.  Drawing from \textit{Emotional Information} theory, which claims that negative sentiment may be a strong indicator of negative links between individuals, we interpret observed negative links as signs of discord or tension (\cite{beigiSocialScienceGuided2020}). This distinction is crucial, as online interactions can be incredibly diverse, representing sentiments ranging from support to hostility. Negative interactions significantly impact offline social networks and individual outcomes (\cite{offerNegativeSocialTies2021}). Signed network representations, which assign positive or negative weights to edges, offer a powerful tool for capturing this complexity.\par 

Previous studies of structural polarization in online media using signed networks have primarily focused on three platforms: Epinions (\cite{richardson2003trust}) and Slashdot (\cite{leskovec2009community}), where users explicitly label each other as friends or foes, and Wikipedia, where users cast votes in administrator elections (\cite{leskovecPredictingPositiveNegative2010}). More recently, signed network representations have been constructed from Reddit and Twitter data, by inferring the interaction sign (positive or negative) using sentiment analysis applied to comments and posts (\cite{pougue2021debagreement, pougue2023learning, keuchenius2021important}). Signed networks allow for a refined analysis of polarization by identifying communities with internal coherence and cross-group antagonism (\cite{harary1953notion, cartwright1956structural, heider1946attitudes, davis1967clustering}). Several methods exist to partition the signed graph into these factions, quantify graph-level polarization, or extract ideological information from the network structure (\cite{aref2019balance, doreian2009partitioning, traag2009community, Kirkley_2019, huang2021pole, arefIdentifyingHiddenCoalitions2021, babul2024sheep}). These studies suggest that negative ties provide a deeper understanding of structural polarization by revealing hidden antagonisms or patterns that may not be apparent in networks only composed of positive interactions (\cite{traag2009community, doreian2009partitioning,arefDetectingCoalitionsOptimally2020, babul2024sheep, keuchenius2021important}). \par

Our work seeks to understand the value of including negative ties through a case study on Menéame, a social media platform with naturally occurring signed signals. Menéame~\footnote{\url{https://www.meneame.net/}} is a popular Spanish social media platform that, with a similar structure to Reddit, primarily functions as a news aggregator. On Menéame, users can post links to news stories; the post appears in Menéame's news feed with the hyperlink, information about the user who posted the story, and a brief description of the story. These stories can then be voted (upvoted or downvoted) or commented on by other users in the ecosystem (\cite{eexploratory}). The articles posted cover a wide range of topics, from sports to local and international politics. Given how users interact on the Menéame platform, we can naturally extract a signed network representation of the user base interactions, where the signed signals can be obtained directly from the up- and downvotes that users can leave on the articles and comments posted by other users.  While Menéame has been studied before from various perspectives (\cite{kaltenbrunner2011comparative, gomezLikelihoodbasedFrameworkAnalysis2013,aragon2017thread}), to the best of our knowledge, the social network of Menéame has never been studied in a way that exploits the natural signed representation of the social network ecosystem and studies polarization in such a context. Unlike other small community datasets, Menéame's data directly captures the sign of interaction dynamics at the comment level (rather than at the user level or inferred sentiment from text). In addition, the platform is designed so that the main newsfeed appears the same for all users, and there is no personalized recommendation algorithm; thus, the presence of structural polarization is due only to user preferences. \par

Our paper contributes to the literature in two ways; the first is that we have collected and made available a dataset from Menéame~\footnote{The Menéame dataset is available at the following link \url{https://doi.org/10.5281/zenodo.15682068}}. Our second contribution to the study of online polarization lies in our assessment of the value of including negative interactions in measuring polarization at the individual level. To do this, we study the patterns of structural polarization by constructing two types of networks from the Menéame data: a signed network representation that includes both positive and negative interactions, and an unsigned network that includes only positive interactions. Prior research on unsigned social media networks has identified structural position at the user level, revealing limited (but existing) communication between opposing groups (\cite{barberaTweetingLeftRight2015a,barberaSocialMediaEcho2020}). Moreover, studies using signed networks primarily seek to identify communities with a high in-group agreement and out-group antagonism (\cite{traag2009community, doreian2009partitioning, keuchenius2021important,esmailianCommunityDetectionSigned2015a,talagaPolarizationMultiscaleStructural2023b}). More recently, methods have been developed to give each user a score that describes how polarized or extreme they are within the context of the network (\cite{babul2024sheep}). To quantify the value of negative ties to the study of polarization, we take a dual approach, leveraging methods designed for both signed and unsigned networks. We use the Signed Hamiltonian Eigenvector Embedding for Proximity (SHEEP) method (\cite{babul2024sheep}) for the signed network, and Correspondence Analysis (CA)  (\cite{greenacre2017correspondence}) for the unsigned network. Both SHEEP and CA produce lower-dimensional embeddings of network data using spectral techniques and matrix decomposition, respectively, and have been used for latent ideology analysis (\cite{babul2024sheep,barberaTweetingLeftRight2015a,flamino_political_2023,falkenberg_growing_2022,falkenbergAffectiveInteractionalPolarization2023,Peralta2024}). By evaluating the two methods side-by-side, we illuminate the insights gained by incorporating negative ties into network models of political polarization.\par
\noindent Our study addresses one key research question: 

\textit{What do negative interactions reveal about polarization that positive interactions cannot?} 

\noindent We find that Menéame users can be grouped into two main ideological factions that exhibit structural polarization. The polarization between factions is much more pronounced in discussions around controversial topics (e.g., the Russia-Ukraine war) compared to discussions of general politics. The two methods we use---SHEEP and Correspondence Analysis (CA)---largely agree in identifying ideological groups, and we verify the ideological groups we detect against an independent ideological measure. CA is perhaps even more effective at identifying disparate ideological groups within the unsigned network. However, negative ties reveal critical patterns, particularly at the extremes, that remain hidden when only positive interactions are analyzed. For instance, in the network of users voting on comments related to the Russia-Ukraine war, the extreme users identified by SHEEP are found to be those who upvote stories from the Russian-state-controlled news outlet \textit{RT} (\textit{Russia Today}), while CA fails to distinguish these users from general left-wing users. More broadly, only SHEEP is capable of detecting extreme users who engage in high levels of antagonism. Our findings suggest that the signed and unsigned networks offer complementary insights and that combining both methodologies enhances our understanding of polarization. 

The paper is organized as follows. The section ``Methods'' describes the data collection process and the construction of the networks and introduces the two techniques we use to analyze polarization: SHEEP for signed networks and Correspondence Analysis for unsigned networks. The section ``Results and Discussion'' presents our findings, comparing the insights gained from the signed and unsigned networks, focusing on topic-specific polarization and the role of the negative interactions. Finally, in the conclusion, we summarize our main contributions, discuss the implications of our findings, and outline potential directions for future research.

%%%%%%%%%%%%%%%%%%%%%%%%%%%%%%%%%%%%%%%%%%%%%%%%%%%%%
% METHODS
%%%%%%%%%%%%%%%%%%%%%%%%%%%%%%%%%%%%%%%%%%%%%%%%%%%%%

\section{Methods}\label{sec:methods}
%%%%%%%%%%%%%%%%%%%%%%%%%%%%%%%%%%%%%%%%%%%%%%%%%%%%%
% Dataset
\subsection{Dataset -- Menéame social media platform}\label{subsec:dataset}
\begin{figure}[ht!]
    \centering
    \includegraphics[width=0.75\linewidth]{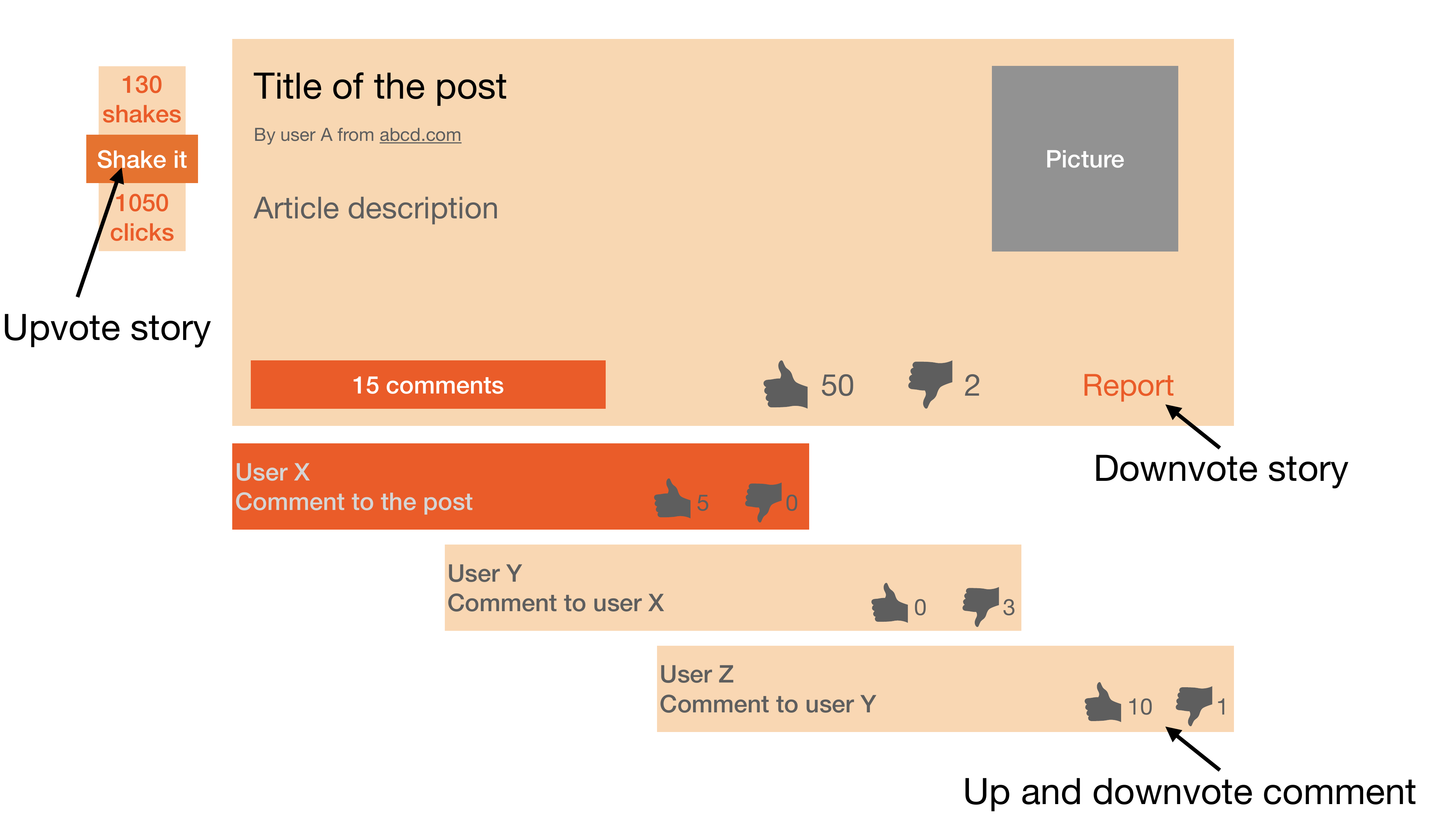}
    \caption{\textbf{Menéame platform}. Schematic representation of one of the stores in the platform. Users can upvote and downvote stories, and upvote and downvote comments within the story. Downvoting stories is possible only for registered users through the ``Report'' button, while upvoting stories is allowed to everyone. Only registered users can vote for comments. Comments with many positive votes appear on the platform highlighted in orange. }
    \label{fig:meneame}
\end{figure}

Menéame is a Spanish news aggregator platform created in 2005 that aims to enhance community participation in information and news diffusion. Users can post and interact with stories, i.e., posts containing a hyperlink to websites such as news outlet articles or social media posts, information about the user who posted the story, and a short description (Fig.~\ref{fig:meneame}). The platform is divided into several sections. We focus on the main page (with the most popular stories) and the queue (where new stories are listed). Users can upvote, downvote, or comment on stories. They can also comment on or vote (up or down) on other comments. Stories that receive the most positive engagement and little negative engagement appear at the top of the feed. 
While the platform guidelines state that users should use negative votes to report spam \footnote{\url{https://www.meneame.net/faq-es}} and to help remove content that goes against the platform guidelines, such uses are rare and result in account removal by moderators. In practice, negative votes are used primarily to express disapproval. 

We collected all the stories, comments, and votes from the period spanning from December 1st, 2022, to August 8th, 2023, comprising 47,887 stories.  In aggregate, our data contains 1,869,190 votes on those stories, 704,636 comments, and 3,113,863 votes on the comments. We found 11,156 unique users voting on stories, while 8,604 users were voting on other users' comments.

In this paper, we create two types of un-directed signed networks: a \textit{user-to-user} network, using the votes on comments; and a \textit{user-to-news outlet} network, using the votes on stories. We create these two networks for different topics, which we extract using the short summaries of the articles. The \textit{user-to-news outlet} network is used to validate our approach since we can compare the ideological position of the news outlet generated by CA and SHEEP to external sources. The \textit{user-to-user} network is our main dataset to study polarization in the platform. Although the format of the data and the platform could naturally produce a directed network representation, we make the modeling choice to remove the directions from the graph to focus on the impact of the signed and weighted edges. Empirically, we found that voting behavior between users tends to be symmetrical, as illustrated in Tables~\ref{tab:crosstab_russia} and~\ref{tab:crosstab_politics}. Details on how we construct the two types of networks can be found in Subsection~\ref{subsec:signet}.

\subsection{Topic Modeling}\label{subsec:topic-modeling}

Topic modeling consists of classifying texts into a finite number of categories (topics) (\cite{bleiTopicModels2009}). It can be supervised if there are existing labels or unsupervised, as in our case, where no topics are indicated in the data. Given the absence of a ``ground truth'' to compare the results of the algorithms, we employ two different algorithms: BERTopic (\cite{grootendorst2022bertopic}) and hierarchical Stochastic Block Model (hSBM) (\cite{gerlachNetworkApproachTopic2018}). We then compare the topics obtained by the two methods, finding robust topics in a restricted subset of stories. In the following, we describe the two algorithms and the comparison technique we implemented.

\subsubsection{BERTopic}\label{subsubsec:bertopic}
The \textit{BERTopic algorithm}, developed by Grootendorst (\cite{grootendorst2022bertopic}), presents a modular text-embedding-based approach for identifying and extracting topics from a given textual dataset based on the BERT (Bidirectional Encoder Representations from Transformers) language model (\cite{devlinBERTPretrainingDeep2019}). The algorithm involves the initial creation of text embeddings, followed by a dimensionality reduction and clustering process applied to these embeddings to form topics. Finally, each topic is associated with keywords using a variation of the Term Frequency-Inverse Document Frequency (TF-IDF). This last step facilitates the interpretation of the topics found with the algorithm. For a thorough description of the algorithm steps and the parameters chosen, please refer to~\ref{appendix:bertopic}.

\subsubsection{hSBM Topic Model}\label{subsubsec:hsbm}
The TM-hSBM, proposed by Gerlach \textit{et al.} (\cite{gerlachNetworkApproachTopic2018}), is an application of community detection methods for topic inference on a text corpus.
In contrast to BERTopic, which is based on embedding sentences into vectors, this approach involves creating a bipartite network with two groups of nodes: words and documents (e.g., comments or stories). Each word is connected to a document if it appears in that document. The method then entails applying a Bayesian hierarchical Stochastic Block Model inference (hSBM) (\cite{hSBM2014}), a method to detect communities in networks, to the bipartite network. In this case, it groups words into topics if they have a unique connectivity pattern---i.e., if they appear and are absent in similar documents (\cite{gerlachNetworkApproachTopic2018}). 
Our detailed procedure is described in~\ref{appendix:hsbm}.

\subsubsection{Combining the results of BERTopic and hSBM}\label{subsubsec:topics}
We apply both BERTopic and TM-hSBM to the story descriptions' text corpus. We found 180 topics using BERTopic containing 38,883 stories and 16,633 stories labeled as outliers. Table~\ref{tab:topics-bertopic} in the Appendix shows the topic names and the number of stories per topic. We performed outlier reduction techniques based on cosine similarities between embeddings. We also reduced the number of topics to 30. 

TM-hSBM provides topics at different hierarchical levels. At the highest granularity level, it finds 114 topics, which are combined into 20, 4, and 1 topics, respectively. Table~\ref{tab:topics-hsbm} in the Appendix shows the topic keywords for each level. Given the results obtained by the two methods independently, we compared them to obtain more robust topics. We computed the accuracy classification score and found that BERTopic with 30 reduced topics is the most similar to the second level of TM-hSBM. 

We analyzed the most representative keywords for each topic identified using BERTopic based on their TF-IDF weights. For instance, Topic 1 included keywords such as \textit{``ukraine, russia, war, russian (male), russian (female), putin, russian (plural), nato, military''}, which we manually labeled as the \textit{``Russia-Ukraine war''} topic. We then assessed the overlap of each topic with the hSBM results.

For example, \textit{hSBM Topic 1} (keywords: \textit{``ukraine, russian (male), war, russia, ukrainian (male), trump, putin, militar, nato, invasion''}) and \textit{hSBM Topic 16} (keywords: \textit{``nuclear, chinese (male), ukraine, russian (male), militar, fire, russia, wagner, china, american''}) were grouped under the same macro-topic \textit{BERTopic Topic 1} as they had an overlap of 52 and 46\% respectively. In cases where multiple BERTopic topics aligned with the same hSBM topic, we merged the BERTopic topics. For instance, the \textit{hSBM Topic 6} (\textit{``rent, healthcare, hospital, union, labor, doctor, strike''}) was matched with the \textit{BERTopic Topics 11} (\textit{``health, strike, doctors, hospitals''}), \textit{14} (\textit{``workers, labor, work, employment''}), and \textit{18} (\textit{``education, students, teachers, schools''}) and manually labeled as \textit{``Public Services''} (see Figure~\ref{fig:confusion-mat} to observe the overlap).

The combination of BERTopic and hSBM resulted in 7 macro-topics: \textit{Broad Politics, Russia-Ukraine War, Public Services, Crime, Climate Change, Cryptocurrencies/tech, Inflation}. We then kept the stories classified in that topic by \textit{both} algorithms and validated our approach by manually labeling a random sample of the comments. Out of 100 comments on which both algorithms agreed on the topic, we agreed with the classification in 93 instances and disagreed in 7. Out of 100 comments on which algorithms disagreed, we agreed with BERTopic in 34 cases, with hSBM in 26 cases, with neither algorithm in 39 cases, and with both algorithms in 1 case. Table~\ref{tab:data_stats} shows the number of stories for each macro-topic. While only a small fraction of the initial corpus is preserved, this refined categorization enables us to understand whether the importance of negative ties to assess polarization is topic-specific. In this paper, we focus on the two largest topics: \textit{Broad Politics} and \textit{Russia-Ukraine}. 

\begin{table}[ht]
\centering
\begin{tabular}{lcccc}
\toprule
Topic & Number of stories & Number of votes & Number of upvotes & Number of downvotes \\
\midrule
Broad Politics & 7,427 & 411,591 & 378,528 & 33,063 \\
Russia-Ukraine War & 2,394 & 77,828 & 64,240 & 13,588 \\
Public services & 1,621 & 85,154 & 81,637 & 3,517 \\
Crime & 1,537 & 73,465 & 69,074 & 4,391 \\
Climate Change & 1,458 & 41,847 & 39,661 & 2,186 \\
Cryptocurrencies/tech & 814 & 26,533 & 24,414 & 2,119 \\
Inflation & 763 & 26,724 & 25,316 & 1,408 \\
\bottomrule
\end{tabular}
\caption{Number of stories, votes, upvotes, and downvotes per macro-topic in the dataset.}\label{tab:data_stats}
\end{table}

%%%%%%%%%%%%%%%%%%%%%%%%%%%%%%%%%%%%%%%%%%%%%%%%%%%%%
% Networks from data
\subsection{Creating networks from data}\label{subsec:signet}
In the previous section, we described how we used the text corpus composed of short textual descriptions of each news story to divide these texts into macro-topics. Here, we seek to explore the interactions between users and stories. Consequently, we construct two networks from our dataset: a network of \textit{user-to-user} interactions and a network of interactions between \textit{users and news outlets}.

In the first case, we consider a network \( G = (U, E) \), where \( U \) is the set of users, and \( E \) is the set of edges. The network can be represented with an adjacency matrix \( A \) of dimension \( N \times N \) (\( N \) is the number of users), where each entry $A_{i,j}$ is the sum of the signed votes of the user \( i \) to the comments of the user \( j \) and the signed votes of the user \( j \) to the comments of the user \( i \). As a result, the network we construct is undirected, and the adjacency matrix \( A \)  is symmetric. If there is no interaction between the two users, or if there are the same number of positive and negative interactions, \( A_{i,j} = 0 \)~\footnote{We assume that the absence of interaction is comparable to a null sum of interactions (number of positive votes equal to the number of negative votes), motivated by a low number of such ``neutral'' interactions, as they entail only 0.74 \% of the dataset.}. The weight is bounded between \( [-n_{ij}, +n_{ij}] \), where \( n_{ij} \) is the number of interactions between \( i \) and \( j \). Specifically, \( A_{i,j} = -n_{ij} \) in case there are only negative votes between the two users, and \( A_{i,j} = +n_{ij} \) if there are only positive votes. We only consider ``active'' users who cast more than 10 votes during the period studied. This is a one-mode (unipartite) network, as all the nodes are of the same type (users).

In the second case, we consider a two-mode (bipartite) network \( B = (V, E') \), where \( V \) is the set of nodes, and \( E' \) is the set of edges. In this case, nodes are of two types, users and news outlets, that form two disjoint sets, which we label \( U \) and \( O \) to represent users and news outlets. The edges connect nodes from one subset to the other only, i.e., \( E' \subseteq U \times O \). We can represent this network with an adjacency matrix that has the shape
\begin{equation*}
A = \begin{pmatrix}
0 & I\\
I^T & 0
\end{pmatrix},
\end{equation*}
where \( I \) is the incidence matrix that has shape \( |U| \times |O| \), and each entry $I_{k,l}$ is the sum of the signed votes of user \( k \) on the stories from the news outlet \( l \). Note that a user can vote on many different stories from a given news outlet, and this information is aggregated in our network. Similarly to the common practice in latent space models (\cite{pooleSpatialModelLegislative1985})---where the vote (positive/negative) is modeled as depending on the difference between latent ideological positioning---we removed the stories that only received positive votes. We found this step to be fundamental in quantifying the impact of controversial stories, and it reflects the fact that stories with only positive votes do not polarize the discussion. 

%%%%%%%%%%%%%%%%%%%%%%%%%%%%%%%%%%%%%%%%%%%%%%%%%%%%%
% SHEEP
\subsection{SHEEP Embedding}\label{subsec:sheep}
\textit{Signed Hamiltonian Eigenvector Embedding for Proximity} (SHEEP), developed by Babul and Lambiotte (\cite{babul2024sheep}), is a spectral embedding method capable of representing proximal information of nodes, using both positive and negative interactions. SHEEP is based on the minimization of the repelling Laplacian (\cite{shiDynamicsSignedNetworks2019}), defined as
\begin{equation}\label{eq:repelling-laplacian}
    L_r = D^+ - A^+ - D^- - A^-,
\end{equation}
where $D^+$ (resp., $D^-$) and $A^+$ (resp., $A^-$) are the degree and adjacency matrix of the positive (resp., negative) part of the network. Babul and Lambiotte (\cite{babul2024sheep}) proved the equivalence between the spectrum of the repelling Laplacian and the Hamiltonian in one dimension, where positive edges are considered as spring attractive forces and negative edges are anti-spring repulsive forces, as follows.

\begin{equation}\label{eq:sheep}{{{{{{{{\boldsymbol{\pi }}}}}}}}}^{T}{L}_{r}{{{{{{{\boldsymbol{\pi }}}}}}}}=\mathop{\sum}\limits_{i,j}{A}_{ij}^{+}| {\pi }_{i}-{\pi }_{j}{| }^{2}+\mathop{\sum}\limits_{i,j}{A}_{ij}^{-}| {\pi }_{i}-{\pi }_{j}{| }^{2}.
\end{equation}

In higher dimensions, the algorithm associates each node in the network with a position (i.e., an embedding $\pi_i \in \mathbb{R}^{N}$), generated using the first $k$ eigenvectors of the repelling Laplacian, such that nodes connected by positive edges are placed closer together, and nodes connected by negative edges are placed further apart. The algorithm also provides a method to identify the optimal dimension for the embedding, by minimizing a generalized version of the Hamiltonian in Eq.~\ref{eq:sheep} (for more details see~\cite{babul2024sheep}). This method, when applied to a signed network of bill co-sponsorship frequency in the US House of Representatives, is successful at recovering the political ideology of the House members on a continuous spectrum (\cite{babul2024sheep}). 

We apply the SHEEP embedding method for each topic to both the unipartite signed network of \textit{user-to-user} votes and the bipartite signed networks of \textit{user-to-news outlet} votes. We use a normalized version of the adjacency matrix, where the entries are divided by the total degree of the network. Following this procedure, we obtain an embedding for each user and news outlet, which we use for further analysis.
Since nodes with a high number of negative votes are pushed away in one of the dimensions from all other nodes, we project each embedding into one dimension using a principal component analysis (PCA) projection. 

%%%%%%%%%%%%%%%%%%%%%%%%%%%%%%%%%%%%%%%%%%%%%%%%%%%%%
% CA
\subsection{Correspondence Analysis}\label{subsec:ca}
\textit{Correspondence Analysis} (CA), first theorized by Hirschfeld (\cite{Hirschfeld_1935}) and later applied by Benzécri et al. (\cite{benzécri1973analyse}), is a widely used method to obtain lower-dimensional representations of data, especially networks.  As described in Greenacre's works (\cite{greenacre1984theory, greenacre2017correspondence}), CA is a statistical technique that produces embeddings of categorical data in a lower-dimensional space. Similar to the principal component analysis, it is based on the singular value decomposition (SVD) of the interaction matrix. For example, in our case, we have a matrix where the rows are the users, the columns are the news outlet domains, and the entries are the number of times each user interacted with the news outlet. In detail, given the interaction matrix $\mathbf{I} \in \mathbb{N}^{a \times b}$, where $a$ is the number of users, in our case, and $b$ is the number of outlets, we first compute the correspondence matrix $\mathbf{P} = \frac{1}{n} \mathbf{I}$, where $n = \sum_i \sum_j I_{ij}$ is the sum of all the entries. Then, we compute the matrix of standardized residuals.

\begin{equation}\label{eq:CA-residuals}
    \mathbf{S} = \mathbf{D}_r^{-\frac{1}{2}} (\mathbf{P} - \mathbf{r} \mathbf{c}^T ) \mathbf{D}_c^{-\frac{1}{2}},      
\end{equation}

where $\mathbf{r} = \mathbf{P} \; \mathbf{1}$ and $\mathbf{r} = \mathbf{P}^T \; \mathbf{1}$, i.e., the so-called row and column masses, and $\mathbf{D}_x = \text{diag}(\mathbf{x})$. Then we calculate the SVD of the matrix $\mathbf{S} = \mathbf{U} \Lambda \mathbf{V}^T$ such that $ \mathbf{U}  \mathbf{U}^T =  \mathbf{V}  \mathbf{V}^T =  \mathbf{1}$. The diagonal matrix $\Lambda$ contains the singular values and is used to determine the embeddings of rows (users) and columns (either users in the \textit{user-to-user} network or news outlets in the \textit{user-to-news outlet} network) following 

\begin{align}\label{eq:ca-embeddings}
    \mathbf{CA}_r &= \mathbf{D}_r^{-\frac{1}{2}} \mathbf{U} \Lambda \notag \\
    \mathbf{CA}_c &= \mathbf{D}_r^{-\frac{1}{2}} \mathbf{V} \Lambda
\end{align}

We use the first dimension of the embeddings recovered in Eqs.~\ref{eq:ca-embeddings} for further analysis. In this work, we employ the \textit{prince} Python package (\cite{Halford_Prince}), which performs CA among other statistical techniques.

%%%%%%%%%%%%%%%%%%%%%%%%%%%%%%%%%%%%%%%%%%%%%%%%%%%%%
% Outlet ideology
\subsection{Ideology of news outlets}\label{subsec:twitter-ideology}
To validate the results in the \textit{user-to-news outlet} network of the two methods described above, we created two independent ``ground truth'' measures of the left-right ideological position of news outlets. First, we calculated ideological positions using Twitter (now X) data.  We used the Tweepy Python package (\cite{harmonTweepy2023}) and the Twitter API v2 with Academic Research access. We collected all tweets from the main Spanish political parties that are influential in terms of popular votes, excluding regionalist parties that primarily tweet in languages other than Spanish. The analyzed parties were \textit{PP, PSOE, CS, PODEMOS, IU, VOX, MasPais, and  PACMA}. Additionally, we identified the 20 most mentioned accounts by each party's account and manually filtered out accounts not associated with politicians or institutions linked to the same party. The complete list of Twitter handles can be found in the \ref{app:twitter}. 

Next, we counted the number of tweets per political party mentioning one of the 40  Spanish news outlets (by website domain) that are more popular on Meneame\footnote{\textit{rtve, abc, elmundo, atresplayer, cope, okdiario, larazon, ondacero, telecinco, vozpopuli, youtube, elespanol, europapress, elconfidencial, telemadrid, cuatro, canalsur, eltorotv, elindependiente, cadenaser, eleconomista, elpais, lavanguardia, esdiario, libertaddigital, 20minutos, elperiodico, lasexta, lavozdegalicia, eldiario, huffingtonpost, facebook, twitch, infolibre, publico, laultimahora, elsaltodiario, gaceta}}. We then applied Correspondence Analysis to the interaction matrix between political parties and news outlets. The resulting CA embedding for political parties aligns with the left-right division in Spain (see Fig.~\ref{fig:app:twitter} in the Appendix). We use the first dimension of the CA embedding as our ``ground truth'' measure of the left-right ideological position of news outlets. This procedure for obtaining ideological positioning is closely linked to latent space models (\cite{barberaTweetingLeftRight2015a}).

As a robustness test, we used the media positioning provided by PoliticalWatch\footnote{\url{https://politicalwatch.es/blog/political-watch-publica-primer-media-bias-chart-espana-2021/}}. The qualitative analysis of PoliticalWatch evaluates 30 media outlets' characteristics, such as the wording and fact-checking standards on a sample of articles to assess their ideological leaning. We found a very high correlation (96\%) between the ideological position of news outlets determined by PoliticalWatch and our method based on Twitter (Fig.~\ref{fig:app:twitter_media} in the Appendix).

The media positioning provides us with two key opportunities. First, it allows us to test the performance of SHEEP and CA on the user-media outlet network. Second, it allows us to validate our analysis on the \textit{user-to-user} network. For example, users classified as left-wing according only to their interaction with other users can be validated by their vote on left-wing stories in Menéame. 

\clearpage
%%%%%%%%%%%%%%%%%%%%%%%%%%%%%%%%%%%%%%%%%%%%%%%%%%%%%
% RESULTS
%%%%%%%%%%%%%%%%%%%%%%%%%%%%%%%%%%%%%%%%%%%%%%%%%%%%%

\section{Results and Discussion}\label{sec:results}
%%%%%%%%%%%%%%%%%%%%%%%%%%%%%%%%%%%%%%%%%%%%%%%%%%%%%
% News outlets results
\subsection{Mapping the ideological landscapes of news outlets}\label{subsec:domains-ideology}
Understanding the ideological positioning of news outlets is crucial in today’s polarized media environment, especially to discern how contentious topics shape public discourse. This ideological positioning is reflected in the \textit{user-to-news outlet} network, where users are connected to a news outlet domain by an edge with an associated weight obtained by aggregating their votes on stories linked to that outlet. Users with an ideological positioning close to the outlet will be more likely to vote positively, while users with an ideological positioning far away from the outlet will be more likely not to vote or vote negatively. Since the ideological positioning might be issue-dependent (e.g., individuals from both the left and right spectrum may support Ukraine in the Ukraine-Russia war), we focus on the two biggest macro-topics on the platform during the specified time frame: \textit{Russia-Ukraine war} and \textit{Broad Politics}. 

We use SHEEP and CA to quantify the structural position of each news outlet and compare these results to external benchmarks of left-right ideology (See Methods Section~\ref{subsec:twitter-ideology}). We find that incorporating negative ties allows us to uncover patterns of structural polarization that would otherwise remain hidden, especially when observing divisive issues, such as the Russia-Ukraine war.

\begin{figure}[ht!]
    \centering
    \includegraphics[width=\linewidth]{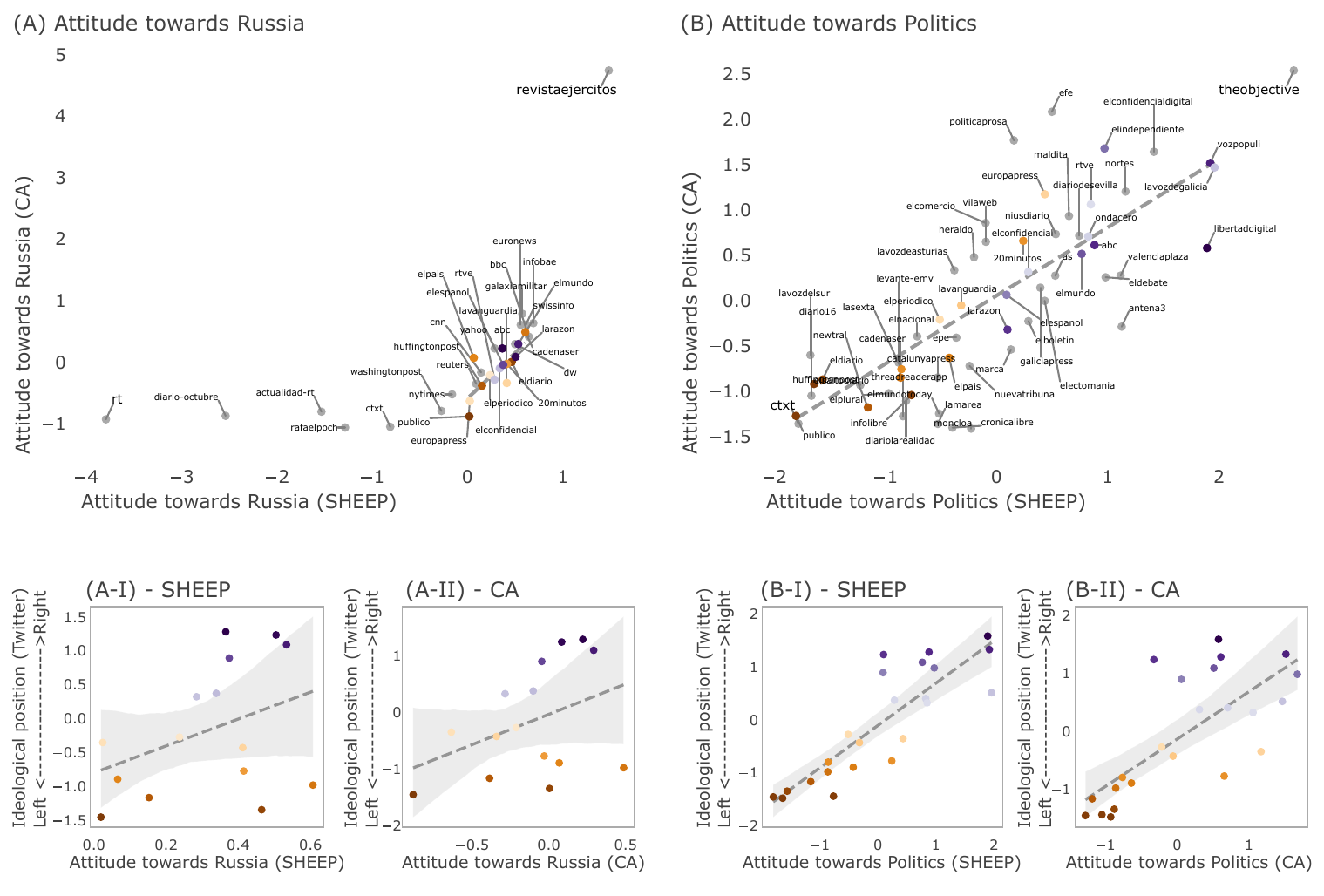}
    \caption{\textbf{Comparing News Outlets' Attitudes towards the Russia-Ukraine War and Politics.} The main panels (A--B) display the embeddings for each news outlet, obtained by both positive and negative ties (SHEEP) and only positive ties (CA). The smaller panels (x-I) and (x-II) compare the two embedding techniques with the ideology retrieved from Twitter for a subset of news outlets. Colors represent the Twitter ideology in all panels, ranging from left-wing (brown) to right-wing (dark purple), while news outlets not classified are colored in grey. In the case of Russia (panel A), both CA and SHEEP identify the army-related news outlet \textit{Revista Ejércitos} as an outlier, whereas only SHEEP distinguishes the pro-Russia news outlets \textit{Russia Today}, \textit{Diario Octubre}, and \textit{Actualidad RT} from the left-leaning media such as \textit{ctxt} or \textit{publico}. Panel B shows that the two methods are highly correlated, both identifying ideology in accordance with Twitter (see panels B-I and B-II) when considering the Politics topic. In contrast, panels A-I and A-II show that both SHEEP and CA are less correlated with Twitter ideology in the case of the Russia-Ukraine topic.
    }
    \label{fig:domains}
\end{figure}

%%%%%%%%%%%%%%%%%%%%%%%%%%%%%%%%%%%%%%%%%%%%%%%%%%%%%
% Russia Ukraine
\subsubsection{Ideological Mapping in the Russia-Ukraine war}\label{subsubsec:domains-russia}
We compare the embeddings of the news outlets nodes obtained by performing SHEEP and CA on the bipartite \textit{user-to-news outlet} network generated by stories classified as belonging to the \textit{Russia-Ukraine war}. We find a moderate Pearson correlation of 58\%  between CA and SHEEP (Figure~\ref{fig:domains}A). The moderate correlation is due to the unique ability of SHEEP to correctly identify \textit{Russia Today (rt)}, \textit{Diario Octubre}, and \textit{Actualidad RT} as extreme in the pro-Russia faction. In contrast, CA maps independent left-wing outlets such as \textit{CTXT} or \textit{Publico} at a similar level of attitude as Russia-funded outlets. This divergence between methods highlights the role of negative ties in distinguishing left-wing positions from pro-Russia. Both methods highlight the news outlet \textit{Revista Ejércitos}---an outlet that, according to their website ``aims to influence political agendas, highlight defense gaps, and promote public investment in Spain's defense industry''---as the most extreme in the other direction. 

We then compared the similarity between the embeddings of SHEEP (Fig.~\ref {fig:domains}A-I) and CA (Fig.~\ref{fig:domains}A-II) with the validated left-right ideology identified from Twitter data (see the Methods section~\ref{subsec:twitter-ideology} for more details). The ideological positioning obtained from SHEEP and CA shows only a moderate Pearson correlation (38\% and 39\%, respectively) with left-right ideology. This finding underscores that, while Twitter-based ideologies derived from the outlets cited by political parties reflect general political leanings, they do not fully capture issue-specific stances in polarized topics. For example, the left-leaning outlets \textit{eldiario}, \textit{20minutos}, and \textit{cadenaser} are found to be strongly opposed to Russia in this context.

%%%%%%%%%%%%%%%%%%%%%%%%%%%%%%%%%%%%%%%%%%%%%%%%%%%%%
% Broad politics
\subsubsection{Ideological Mapping in Broad Politics}\label{subsubsec:domains-politics}
%he correlation was notably higher at 76% 
After finding that negative ties allow us to uniquely find pro-Russia outlets, we looked at the embeddings of news outlets on the \textit{user-to-news outlet} network generated from stories in the ``Broad Politics'' topic, which is a macro-topic incorporating stories about Spanish politics. We find a high Pearson correlation (76\%) between the embeddings generated by CA and SHEEP (Fig.~\ref{fig:domains}B). The results are also highly correlated (86.6\% for SHEEP and 73.1\% for CA)  with the ideology retrieved from Twitter (Figures~\ref{fig:domains}B-I and B-II). 

These results suggest that when applied to the broad politics topic, both methods produce results that align with the left-right political spectrum, but on the Russia–Ukraine War, the embeddings diverge, reflecting issue-specific dimensions that are captured by negative ties.

%%%%%%%%%%%%%%%%%%%%%%%%%%%%%%%%%%%%%%%%%%%%%%%%%%%%%
% User level
\subsection{Examining Structural Polarization at the User Level}\label{subsec:users-ideology}
This section brings us closer to the central question of the paper: \textit{what do negative interactions---downvotes---reveal about polarization that remains hidden when we consider only positive engagements?} We estimate the SHEEP and CA embeddings for the users using the unipartite \textit{user-to-user} networks, where users vote on the comments of other users. We consider the same two topics as in the analysis at the outlet level: \textit{Russia-Ukraine War} and \textit{Broad Politics}.

%%%%%%%%%%%%%%%%%%%%%%%%%%%%%%%%%%%%%%%%%%%%%%%%%%%%%
% Russia-Ukraine
\subsubsection{Polarized Factions and Antagonism in the Russia-Ukraine  Topic}\label{subsubsec:users-russia}

\begin{figure}
    \centering
    \includegraphics[width=1\linewidth]{figures/fig3.pdf}
    \caption{\textbf{Visualization of \textit{user-to-user} network for the Russia-Ukraine war topic.} Both panels share the same layout, generated using the Fruchterman-Reingold force-directed algorithm, but not all the nodes appear on both networks. A random sample of 3,000 nodes is shown. Edges represent interactions: positive in blue, negative in red (only in SHEEP), and those with absolute weight smaller than 3 are filtered out. Node colors show standardized SHEEP (A) and CA (B) embedding values, capped at 2, with red indicating anti-NATO and blue pro-NATO attitudes. The layout reveals two ideological factions with notable cross-faction interaction. Note that for SHEEP (A), the most extreme users (darker shades of blue and red) are located within the network, while for CA (B), they tend to be the ones with a few votes and thus no visible edges.
    Panel C compares SHEEP and CA in determining user attitudes: each point is a user, with the color indicating their tendency to vote positively (dark blue) or negatively (red). Blue and red circles highlight users voting positively on \textit{Revista Ejércitos} and \textit{Russia Today}, respectively. Histograms show user distribution across the embedding space for each method. Note that SHEEP identifies extreme negative voters in both factions, and the methods are strongly correlated, with a Spearman correlation of 88\%.}    
    \label{fig:network_viz_russia}
\end{figure}

We begin by visualizing the network using the Fruchterman-Reingold force-directed algorithm (\cite{fruchtermanGraphDrawingForcedirected1991}), considering only positive interactions (Fig. \ref{fig:network_viz_russia}). This algorithm (often called Spring Layout) works by using repulsion (between non-connected nodes) and attraction (between connected nodes with positive interactions) to position nodes in two dimensions. Both the visualization produced by Spring Layout and the embeddings generated by SHEEP and CA (indicated by node colors in Fig.~\ref{fig:network_viz_russia}A-B) reveal two distinct factions. By manually reading the comments, we identify these factions as against NATO and pro-arming Ukraine~\footnote{An example comment of each side is ``You speak as if Russia is preventing Ukraine from joining UNICEF. NATO is not just any ``supranational body'', it is a military alliance with its missiles pointed at Russia.'' and ``Letting Russia do whatever the hell it wants is much more dangerous than arming Ukraine. Among other things because otherwise in a year you would have 3 or 4 other major countries taking example and going over the top of international laws knowing that they have more legitimacy than before.''}. While SHEEP and CA demonstrate significant agreement in the overall classification of users within these factions, they differ significantly in identifying the most extreme individuals. SHEEP pulls disengaged users (those with few votes) toward the center of the embedding, while this pattern is less pronounced for CA---the Spearman correlations between degree centrality and SHEEP and CA embeddings are 36\% and 22\% respectively. As a result, SHEEP identifies extreme users as those with a more active, hostile behaviour, i.e., with more negative links (dark red nodes in Fig.~\ref{fig:network_viz_russia}A).

Examining the similarities and differences between SHEEP and CA, we find that the embeddings generated by both methods are highly correlated (Fig.~\ref{fig:network_viz_russia}C), with a Spearman correlation of 88\%. As observed in the network visualization, SHEEP and CA only disagree in identifying the most extreme users. In the context of the \textit{Russia-Ukraine war}, both SHEEP and CA exhibit bimodal distributions (histograms in Fig.~\ref{fig:network_viz_russia}C), which correspond to two primary ideological factions on Menéame: pro-NATO users and anti-NATO users. Both methods identify a larger prevalence of users with a negative stance towards NATO. While both methods exhibit a bimodal distribution, indicative of polarization, the SHEEP embedding displays a long tail at both extremes, while CA does not. The differences in the tails are driven by negative votes. This is evident in the color coding in Fig.~\ref{fig:network_viz_russia}C, where red indicates users who predominantly cast negative votes, and blue indicates those who predominantly vote positively. Users located at the tail-ends of the SHEEP embedding are typically associated with negative voting behavior and frequently interact with the two news outlets identified as polarizing: \textit{RT} and \textit{RevistaEjércitos}. Users who vote positively for these outlets are respectively marked with red and blue circles in Fig.~\ref{fig:network_viz_russia}C. As a result, CA often fails to distinguish extremist pro-Russia users from left-wing users who criticize the role of NATO in the years preceding the Russian invasion. 

To understand to what extent SHEEP and CA can recover ideological factions with homogeneous voting behavior, we analyze the normalized voting probability (Fig.~\ref{fig:users-russia}C--D) and average vote sign (Fig.~\ref{fig:users-russia}E--F) as a function of the embedding created by each method. These matrices are used to identify ideological factions using the k-means clustering method (See~\ref{app:sec:clustering} for more details) and to facilitate the interpretation of the results. We then examined the voting propensity of the three factions (anti-NATO, moderate, pro-NATO) identified by both SHEEP (boxes in Fig.~\ref{fig:users-russia}E and histograms) and CA (boxes in Fig.~\ref{fig:users-russia}F and histograms). 

The majority of the votes of each faction are made to others in the same faction (within the boxes in Figs.~\ref{fig:users-russia}C--D). Interestingly, SHEEP identifies a larger ``moderate'' faction. This is due to the difference in the mapping of users with few votes. While CA considers users with a few votes to extreme users to be themselves extreme, SHEEP considers these users as moderate, as there is not enough information about their latent ideology to pull them far away from the center (see also Fig.~\ref{fig:network_viz_russia}). 

The voting propensities match well with the sign of votes cast by users. The pro and anti-NATO factions vote positively for users with similar attitudes and negatively for users with different attitudes (Figures~\ref{fig:users-russia}E--F). Moreover, while the moderates do not vote negatively for either the pro or anti-NATO factions, they only vote positively for the anti-NATO faction, indicating a closer ideological affinity with this group. Interestingly, this pattern is also found for the CA method, in which negative ties are excluded. This indicates that the information encoded by negative ties is partially available in positive ties---i.e., the absence of interaction is related to the propensity to vote negatively. 

%Since a user can vote either positive (+1) or negative (-1), the average vote ranges between -1 and +1. 
%Noting that only SHEEP accounts for negative votes in the determination of user attitudes, we find that users overall tend to vote positively. 
 %As expected, the lowest average vote is found between the extreme factions identified by both methods.

%Additionally, we find three clusters in the voting pattern matrices, distinguished as colors in the histograms and as squares in Figures~\ref{fig:users-russia}B-C-D-E. The three clusters represent left-center-right divisions, without showing particular distinction for more extreme pro/against Russia users, except for SHEEP's histogram, which shows a peak of pro-Russia users further from the other left-leaning users. We demonstrate that the three clusters exhibit similar patterns in both methods, with a high propensity for and more positive votes within their respective clusters.

Both CA and SHEEP can identify the ideological polarization on the Russia-Ukraine war on the platform, in terms of the bimodal distribution in the embeddings, while only SHEEP describes the inter-faction hostility that we observe as negative voting patterns between extreme users in opposing factions.  

%%%%%%%%%%%%%%%%%%%%%%%%%%%%%%%%%%%%%%%%%%%%%%%%%%%%%
% Politics
\subsubsection{Polarized Factions and Antagonism in the Broad Politics Topic}\label{subsubsec:users-politics}

\begin{figure}
    \centering
    \includegraphics[width=1\linewidth]{figures/fig4.pdf}
    \caption{\textbf{Visualization of \textit{user-to-user} network for the Broad Politics.} Both panels share the same layout, generated using the Fruchterman-Reingold force-directed algorithm. A random sample of 3,000 nodes is shown. Edges represent interactions: positive in blue, negative in red (only in SHEEP), and those with absolute weight smaller than 5 are filtered out. Node colors show standardized SHEEP (A) and CA (B) embedding values, capped at 2. The layout reveals two ideological factions with notable cross-faction interaction. Note that for SHEEP (A), the most extreme users (darker shades of blue and red) are located within the network, while for CA (B), the most extreme users are those with few votes (and thus have no edges visible).   Panel C compares SHEEP and CA in determining user attitudes towards politics. Each point represents a user, with color indicating their tendency to vote positively (dark blue) or negatively (red). Red circles are users who vote positively for \textit{far-left media}. The histograms show the distribution of users across the embedding space for each method.  We note that SHEEP places extremely negative voters in the left faction, while in general, the two methods are consistent in their identification of most users, with a Spearman correlation of 80\%.}
    \label{fig:network_viz_politics}
    \end{figure}

We perform the same analysis as in the previous section, using the \textit{user-to-user} network representing the politics topic. Unlike in the case of the Russia-Ukraine topic network, the visualization of the politics network does not show two clear ideological factions (Fig.~\ref{fig:network_viz_politics}). Instead, we see a more continuous transition from left-wing to right-wing users~\footnote{Two representative comments of extreme users are ``of course, you talk like ``left-wing voters,'' but in the end, you end up saying that you'll let ``the right'' win. Very logical, all of it. Then it's four years of eating shit.''  and ``Isn't that exactly what Sánchez [Spanish PM] wants to do to govern? Despite having fewer votes than the PP [main left-wing party], it seems he'll be the one governing. I haven't seen any criticism about that here... Could it be that everything the left does seems fine to you, and everything the right does seems wrong?''}. As in the case of the Russia-Ukraine network, SHEEP tends to pull active hostile users towards the extremes (i.e., with more negative links (dark red nodes in Fig.~\ref{fig:network_viz_politics}A), resulting in a higher correlation between degree centrality and ideological position (9\% for SHEEP versus 4\% for CA). 

In the case of the broad politics topic, we find more significant differences between the embeddings created by SHEEP and CA (Figs.~\ref{fig:network_viz_politics}C). In the Russia-Ukraine topic, both embeddings display a bimodal distribution. However, in the broad politics topic, only the CA embedding is bimodal, dominated by a large majority of left-wing users. In contrast, the SHEEP embedding exhibits a unimodal distribution with long tails, particularly extending toward far-left users (Fig.~\ref{fig:users-politics}A). While the user distributions in the SHEEP and CA embeddings have different shapes, the position of users within the embedding is remarkably similar (Spearman correlation of 80\%). The main differences occur at the tails, where the methods identify different users as extreme. As in the Russia-Ukraine case, the difference is created by negative votes. Only SHEEP can separate far-left users from other left users (the red circles in Fig.~\ref{fig:network_viz_politics}C indicate users who vote positively towards far-left media).

Performing a k-means clustering on the matrices (see~\ref{app:sec:clustering} for details) of votes and signs identifies four factions of users: far-left, left-leaning, right-leaning, and far-right. 
Users on the left and the right are more likely to vote on comments of users with a similar structural position in the network(Fig.~\ref{fig:users-politics}C--D), but only far-left and far-right users vote negatively towards the opposite faction (Fig.~\ref{fig:users-politics}E--F). 
Interestingly, the most extreme far-left users also exhibit a high propensity to vote negatively to far-right users (bottom-right corner of Figure~\ref{fig:users-politics}E--F), but the opposite is not true: far-right users do not engage as significantly with far-left users. This is likely a platform-specific effect, resulting from the asymmetric distribution of left and right-wing users. 

In summary, our analysis reveals that the platform's audience skews toward left-leaning users. In general, users engage positively with others who share similar views, while far-left extremists use negative votes strategically to target the opposing extreme faction.  We find that negative interactions are necessary to detect these extreme users.

\begin{figure}
    \centering
    \includegraphics[width=0.9\linewidth]{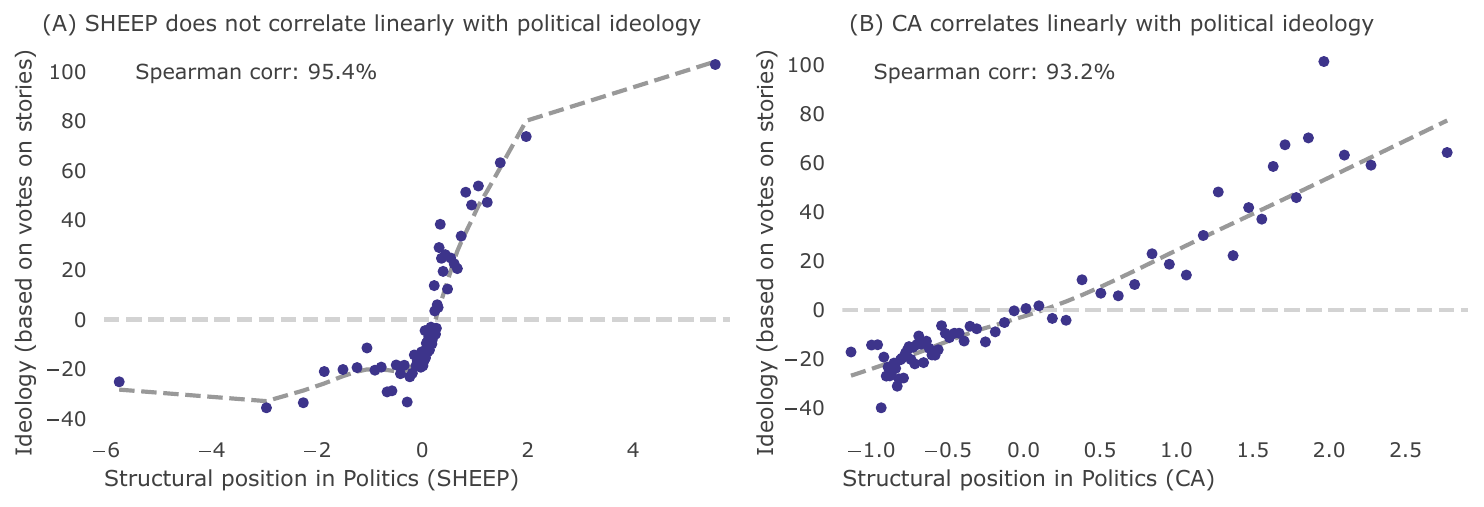}
    \caption{Comparison of the structural positioning derived from two embedding methods: (A) SHEEP and (B) Correspondence Analysis (CA), with ideological positioning. The x-axis represents the binned structural positioning obtained from the embeddings, while the y-axis indicates the average vote of users in each bin toward stories (using the bipartite network described in the previous section). Votes are weighted by the average number of votes per domain and the sign of those votes (see Methods). Points represent ideological bins, with a dashed gray line showing a smoothed regression (lowess) to highlight trends. Positive y-values indicate a higher propensity for and positivity in voting for right-wing news outlets, or a lower propensity for and positivity in voting for left-wing news outlets. Note that CA captures political ideology more linearly compared to SHEEP, which exhibits a non-linear pattern.}
    \label{fig:validation}
\end{figure}

We finally examine whether the structural positions generated by SHEEP and CA align with left-right ideological positions. By comparing the uni-partite \textit{user-to-user} network embeddings to estimates of ideological positioning (see Fig.~\ref{fig:validation}) obtained using the user-domain bipartite network, we found that CA exhibits a linear correlation with left-right political ideology, effectively placing users along the ideological spectrum. Conversely, the SHEEP embedding demonstrates a non-linear relationship with ideology, suggesting that SHEEP captures additional dimensions beyond ideological alignment.

Given that extreme users identified by SHEEP use negative votes more frequently than other users, and that negative interactions may signal emotional hostility (\cite{beigiSocialScienceGuided2020}), SHEEP likely captures underlying elements of this hostility. This phenomenon is particularly pronounced in highly polarizing topics, such as the Russia-Ukraine war, while appearing less pronounced in broader political discussions--except among certain extreme users.

Methodologically, we note that SHEEP and CA take as input two different networks; SHEEP takes a signed network, while CA takes an unsigned network. To confirm that the differences we find using the two methods in the following analysis is indeed due to the addition of the negative ties, rather than the choice of algorithm, we reproduce the above analysis comparing SHEEP and CA, where SHEEP takes as input a null model in which we preserve the existing positive ties and add artificial negative ties to all missing edges, simulating for SHEEP what CA already assumes: that not having a connection implies some level of dissimilarity. We find for both topics that the embeddings generated by the SHEEP null model are highly correlated with those from CA, far more than with the embeddings generated by the original version of SHEEP (see \ref{app:sheep-null}). These results confirm that the differences we observe between the signed and unsigned networks are not artifacts of the algorithm choice but are indeed coming from the addition of negative ties, which offer a more nuanced understanding of the platform polarization.

Overall, the two methods reveal different aspects of both topics. Although negative votes make up only 3\% of all votes on the platform, they are highly relevant for detecting antagonistic hostility in extremely polarized subpopulations--e.g., pro-Russia communities that are conflated with the anti-NATO general faction. Extremist users, both right- or left-leaning, often use negative votes as a tool against the other faction, while more moderate users show support for their ideological camp.

%%%%%%%%%%%%%%%%%%%%%%%%%%%%%%%%%%%%%%%%%%%%%%%%%%%%%
% CONCLUSION
%%%%%%%%%%%%%%%%%%%%%%%%%%%%%%%%%%%%%%%%%%%%%%%%%%%%%

\section{Conclusion}\label{sec:conclusion}
As social media becomes one of the primary mechanisms for ideological exchange and emotional expression, it is an increasingly pressing challenge to understand and quantify polarization in digital environments. This study addresses a key gap by exploring what negative ties reveal about polarization that positive interactions alone cannot. Using data from Menéame, a Spanish social media platform, we investigated the dynamics of polarization through a dual-method approach and created a publicly available dataset for future research (available at \url{https://doi.org/10.5281/zenodo.15682068}).

By comparing the results of SHEEP on the signed network and Correspondence Analysis (CA) on the unsigned network, we identified both ideological divisions and critical interaction patterns, such as antagonism between extreme factions. Both CA and SHEEP successfully map users and outlets along a political spectrum (verified against independent ideology measures from Twitter and PoliticalWatch), with CA somewhat more effectively mapping ideological polarization among users. However, by including negative links through SHEEP, we reveal behaviors that remain hidden in unsigned networks. For example, in the Russia-Ukraine topic, we detect extreme users who are not only involved in antagonistic interactions but consistently upvote content from Russia-affiliated outlets like RT. This highlights the value of negative ties for identifying users who may play a disproportionate role in spreading polarizing or foreign-influenced narratives. 

Our findings show that negative ties are crucial for identifying extreme users who engage in antagonistic behavior. Moreover, we find that far-left users on Menéame are more likely to interact across ideological lines through negative votes, as compared to far-right users, who tend to remain isolated in their interaction patterns. In general, the dual-method approach we take here indicates that the signed and unsigned networks provide complementary insights into the structure of the polarization, and that both can be used together to better understand ideological dynamics—revealing not only users' ideological stances, but how they engage with opposing views and contribute to antagonistic activity.

While the methods we employ here can be applied to any signed social network, some aspects of the analysis may not generalize to other online platforms. One limitation is that our findings are based on a platform with a predominantly left-leaning user base, which may affect the representativeness of the results. Oversampling right-leaning users is not straightforward, as ideological scores come from the embeddings themselves, and oversampling users who frequently upvote right-leaning outlets would also require additional assumptions. We see this as an interesting direction for future work.
More broadly, future research could replicate this framework on other online platforms with negative interactions to allow for comparisons of user behavior and political leanings. The methods we present could also be modified to combine both CA and SHEEP into one embedding, to allow for an ``interpolation'' between the information we obtain from each, which could be a rich area for future investigation.  

Our study advances the theoretical understanding of online polarization, offering methods to identify extreme users and their behaviors. Our conclusions could inform strategies for mitigating the negative effects of online social media and fostering healthier, more constructive online conversations. Ultimately, this work underscores the complex interplay between ideology, emotion, and interaction in digital spaces, contributing to the broader literature on polarization in online networks and methodologies for studying signed interaction data.

\paragraph{Acknowledgments}
We are grateful to R. Lambiotte, J. Pougué-Biyong, and Dorian Quelle for useful conversations and feedback on this work. 

\paragraph{Funding Statement}
S.A. Babul was supported by EPSRC grant EP/W523781/1, and The Alan Turing Institute’s Enrichment Scheme 2023-2024.  J. Garcia-Bernardo acknowledges support from the Dutch Research Council (NWO, grant VI.Veni.231S.148).

\paragraph{Competing Interests}
None.

\paragraph{Data Availability Statement}
The code associated with this research is publicly available at the following URL: \url{https://doi.org/10.5281/zenodo.15496623}. This repository includes the code implementations of the methods discussed in the paper. The dataset is also available on Zenodo at the URL: \url{https://doi.org/10.5281/zenodo.15682068}. 

\paragraph{Ethical Statement}
The data was collected using a custom scraper of 
\texttt{meneame.net}. Ethical approval for this study was obtained from the Faculty Ethics Review Board (FERB) of Utrecht University (Approval number 23-0111). The data published in this article does not include usernames and has been aggregated to avoid possible identification.

\paragraph{Author Contributions}
Conceptualization: JGB, SAB, AB, EC; Data curation: JGB, EC; Formal analysis: SAB, EC, JGB; Methodology: all; Supervision: JGB, AB; Visualization: JGB, EC; Writing–original draft: SAB, EC; Writing–review \& editing: all.

\paragraph{Declaration of generative AI and AI-assisted technologies in the writing process}
During the preparation of this work, the author(s) used https://www.deepl.com/write and ChatGPT 4o for copy-
editing and to improve readability. After using this tool/service, the authors reviewed and edited the content as needed and take(s) full responsibility for the content of the publication.

\bibliography{bibliography}

\begin{thebibliography}{10}

\bibitem{marchi2012facebook}
Regina Marchi.
\newblock With facebook, blogs, and fake news, teens reject journalistic “objectivity”.
\newblock {\em Journal of communication inquiry}, 36(3):246--262, 2012.

\bibitem{garimella2017reducing}
Kiran Garimella, Gianmarco De~Francisci~Morales, Aristides Gionis, and Michael Mathioudakis.
\newblock Reducing controversy by connecting opposing views.
\newblock In {\em Proceedings of the tenth ACM international conference on web search and data mining}, pages 81--90, 2017.

\bibitem{adamic2005political}
Lada~A Adamic and Natalie Glance.
\newblock The political blogosphere and the 2004 us election: divided they blog.
\newblock In {\em Proceedings of the 3rd international workshop on Link discovery}, pages 36--43, 2005.

\bibitem{conover2011political}
Michael Conover, Jacob Ratkiewicz, Matthew Francisco, Bruno Gon{\c{c}}alves, Filippo Menczer, and Alessandro Flammini.
\newblock Political polarization on twitter.
\newblock In {\em Proceedings of the international aaai conference on web and social media}, volume~5, pages 89--96, 2011.

\bibitem{hohmannQuantifyingIdeologicalPolarization2023a}
Marilena Hohmann, Karel Devriendt, and Michele Coscia.
\newblock Quantifying ideological polarization on a network using generalized {{Euclidean}} distance.
\newblock {\em Science Advances}, 9(9):eabq2044, March 2023.

\bibitem{barberaSocialMediaEcho2020}
Pablo Barber{\'a}.
\newblock Social {{Media}}, {{Echo Chambers}}, and {{Political Polarization}}.
\newblock In Joshua~A. Tucker and Nathaniel Persily, editors, {\em Social {{Media}} and {{Democracy}}}, {{SSRC Anxieties}} of {{Democracy}}, pages 34--55. Cambridge University Press, Cambridge, 2020.

\bibitem{falkenberg_growing_2022}
Max Falkenberg, Alessandro Galeazzi, Maddalena Torricelli, Niccolò Di~Marco, Francesca Larosa, Madalina Sas, Amin Mekacher, Warren Pearce, Fabiana Zollo, Walter Quattrociocchi, and Andrea Baronchelli.
\newblock Growing polarization around climate change on social media.
\newblock {\em Nature Climate Change}, 12(12):1114--1121, December 2022.

\bibitem{flamino_political_2023}
James Flamino, Alessandro Galeazzi, Stuart Feldman, Michael~W. Macy, Brendan Cross, Zhenkun Zhou, Matteo Serafino, Alexandre Bovet, Hernán~A. Makse, and Boleslaw~K. Szymanski.
\newblock Political polarization of news media and influencers on {Twitter} in the 2016 and 2020 {US} presidential elections.
\newblock {\em Nature Human Behaviour}, 7(6):904--916, March 2023.

\bibitem{tornberg2022digital}
Petter T{\"o}rnberg.
\newblock How digital media drive affective polarization through partisan sorting.
\newblock {\em Proceedings of the National Academy of Sciences}, 119(42):e2207159119, 2022.

\bibitem{nealSignTimesWeak2020a}
Zachary~P. Neal.
\newblock A sign of the times? {{Weak}} and strong polarization in the {{US Congress}}, 1973--2016.
\newblock {\em Social networks}, 60:103--112, 2020.

\bibitem{abramowitzPolarizationMyth2008}
Alan~I. Abramowitz and Kyle~L. Saunders.
\newblock Is {{Polarization}} a {{Myth}}?
\newblock {\em The Journal of Politics}, 70(2):542--555, April 2008.

\bibitem{abramowitzWhyCantWe2005}
Alan Abramowitz and Kyle Saunders.
\newblock Why {{Can}}'t {{We All Just Get Along}}? {{The Reality}} of a {{Polarized America}}.
\newblock {\em The Forum}, 3(2), July 2005.

\bibitem{dimaggioHaveAmericansSocial1996}
Paul DiMaggio, John Evans, and Bethany Bryson.
\newblock Have {{Americans}}' social attitudes become more polarized?
\newblock {\em American Journal of Sociology}, 102(3):690--755, 1996.

\bibitem{bramsonUnderstandingPolarizationMeanings2017}
Aaron Bramson, Patrick Grim, Daniel~J. Singer, William~J. Berger, Graham Sack, Steven Fisher, Carissa Flocken, and Bennett Holman.
\newblock Understanding polarization: {{Meanings}}, measures, and model evaluation.
\newblock {\em Philosophy of Science}, 84(1):115--159, January 2017.

\bibitem{lelkes2016mass}
Yphtach Lelkes.
\newblock Mass polarization: Manifestations and measurements.
\newblock {\em Public Opinion Quarterly}, 80(S1):392--410, 2016.

\bibitem{barberaTweetingLeftRight2015a}
Pablo Barber{\'a}, John~T. Jost, Jonathan Nagler, Joshua~A. Tucker, and Richard Bonneau.
\newblock Tweeting {{From Left}} to {{Right}}: {{Is Online Political Communication More Than}} an {{Echo Chamber}}?
\newblock {\em Psychological Science}, 26(10):1531--1542, October 2015.

\bibitem{esauDestructivePolarizationDigital}
Katharina Esau, Tariq Choucair, Samantha Vilkins, Sebastian F.~K. Svegaard, Axel Bruns, Kate~S. {O'Connor-Farfan}, and Carly {Lubicz-Zaorski}.
\newblock Destructive polarization in digital communication contexts: A critical review and conceptual framework.
\newblock {\em Information, Communication \& Society}, 0(0):1--22.

\bibitem{salloumSeparatingPolarizationNoise2021}
Ali Salloum, Ted Hsuan~Yun Chen, and Mikko Kivel{\"a}.
\newblock Separating {{Polarization}} from {{Noise}}: {{Comparison}} and {{Normalization}} of {{Structural Polarization Measures}}, December 2021.

\bibitem{yarchiPoliticalPolarizationDigital2020}
Moran Yarchi, Christian Baden, and Neta {Kligler-Vilenchik}.
\newblock Political {{Polarization}} on the {{Digital Sphere}}: {{A Cross-platform}}, {{Over-time Analysis}} of {{Interactional}}, {{Positional}}, and {{Affective Polarization}} on {{Social Media}}.
\newblock {\em Political Communication}, 38:1--42, July 2020.

\bibitem{falkenbergAffectiveInteractionalPolarization2023}
Max Falkenberg, Fabiana Zollo, Walter Quattrociocchi, J{\"u}rgen Pfeffer, and Andrea Baronchelli.
\newblock Affective and interactional polarization align across countries, November 2023.

\bibitem{delvicarioSpreadingMisinformationOnline2016}
Michela Del~Vicario, Alessandro Bessi, Fabiana Zollo, Fabio Petroni, Antonio Scala, Guido Caldarelli, H.~Eugene Stanley, and Walter Quattrociocchi.
\newblock The spreading of misinformation online.
\newblock {\em Proceedings of the National Academy of Sciences}, 113(3):554--559, January 2016.

\bibitem{cinelliEchoChamberEffect2021a}
Matteo Cinelli, Gianmarco De~Francisci~Morales, Alessandro Galeazzi, Walter Quattrociocchi, and Michele Starnini.
\newblock The echo chamber effect on social media.
\newblock {\em Proceedings of the National Academy of Sciences}, 118(9):e2023301118, March 2021.

\bibitem{ferrazdearrudaModellingHowSocial2022a}
Henrique {Ferraz de Arruda}, Felipe Maciel~Cardoso, Guilherme {Ferraz de Arruda}, Alexis R.~Hern{\'a}ndez, Luciano {da Fontoura Costa}, and Yamir Moreno.
\newblock Modelling how social network algorithms can influence opinion polarization.
\newblock {\em Information Sciences}, 588:265--278, April 2022.

\bibitem{ribeiro2017everything}
Manoel~Horta Ribeiro, Pedro~H Calais, Virg{\'\i}lio~AF Almeida, and Wagner Meira~Jr.
\newblock " everything i disagree with is\# fakenews": Correlating political polarization and spread of misinformation.
\newblock {\em arXiv preprint arXiv:1706.05924}, 2017.

\bibitem{enelowSpatialTheoryVoting1984}
James~M. Enelow and Melvin~J. Hinich.
\newblock {\em The {{Spatial Theory}} of {{Voting}}: {{An Introduction}}}.
\newblock CUP Archive, April 1984.

\bibitem{pooleSpatialModelLegislative1985}
Keith~T. Poole and Howard Rosenthal.
\newblock A {{Spatial Model}} for {{Legislative Roll Call Analysis}}.
\newblock {\em American Journal of Political Science}, 29(2):357--384, 1985.

\bibitem{pooleSpatialModelsParliamentary2005}
Keith~T. Poole.
\newblock {\em Spatial Models of Parliamentary Voting}.
\newblock Analytical Methods for Social Research. Cambridge University Press, Cambridge ; New York, 2005.

\bibitem{clintonStatisticalAnalysisRoll2004}
Joshua Clinton, Simon Jackman, and Douglas Rivers.
\newblock The {{Statistical Analysis}} of {{Roll Call Data}}.
\newblock {\em American Political Science Review}, 98(2):355--370, May 2004.

\bibitem{yuSpatialVotingModels2021}
Xingchen Yu and Abel Rodr{\'i}guez.
\newblock Spatial voting models in circular spaces: {{A}} case study of the {{U}}.{{S}}. {{House}} of {{Representatives}}.
\newblock {\em The Annals of Applied Statistics}, 15(4):1897--1922, December 2021.

\bibitem{moodyPortraitPoliticalParty2013}
James Moody and Peter~J. Mucha.
\newblock Portrait of {{Political Party Polarization}}.
\newblock {\em Network Science}, 1(1):119--121, April 2013.

\bibitem{waughPartyPolarizationCongress2009}
Andrew~Scott Waugh, Liuyi Pei, James~H. Fowler, Peter~J. Mucha, and Mason~A. Porter.
\newblock Party {{Polarization}} in {{Congress}}: {{A Network Science Approach}}.
\newblock 2009.

\bibitem{barberaBirdsSameFeather2015}
Pablo Barber{\'a}.
\newblock Birds of the {{Same Feather Tweet Together}}: {{Bayesian Ideal Point Estimation Using Twitter Data}}.
\newblock {\em Political Analysis}, 23(1):76--91, January 2015.

\bibitem{beigiSocialScienceGuided2020}
Ghazaleh Beigi, Jiliang Tang, and Huan Liu.
\newblock Social {{Science}}--guided {{Feature Engineering}}: {{A Novel Approach}} to {{Signed Link Analysis}}.
\newblock {\em ACM Transactions on Intelligent Systems and Technology}, 11(1):1--27, February 2020.

\bibitem{offerNegativeSocialTies2021}
Shira Offer.
\newblock Negative {{Social Ties}}: {{Prevalence}} and {{Consequences}}.
\newblock {\em Annual Review of Sociology}, 47(Volume 47, 2021):177--196, July 2021.

\bibitem{richardson2003trust}
Matthew Richardson, Rakesh Agrawal, and Pedro Domingos.
\newblock Trust management for the semantic web.
\newblock In {\em The Semantic Web-ISWC 2003: Second International Semantic Web Conference, Sanibel Island, FL, USA, October 20-23, 2003. Proceedings 2}, pages 351--368. Springer, 2003.

\bibitem{leskovec2009community}
Jure Leskovec, Kevin~J Lang, Anirban Dasgupta, and Michael~W Mahoney.
\newblock Community structure in large networks: Natural cluster sizes and the absence of large well-defined clusters.
\newblock {\em Internet Mathematics}, 6(1):29--123, 2009.

\bibitem{leskovecPredictingPositiveNegative2010}
Jure Leskovec, Daniel Huttenlocher, and Jon Kleinberg.
\newblock Predicting positive and negative links in online social networks.
\newblock In {\em Proceedings of the 19th International Conference on {{World}} Wide Web}, pages 641--650, Raleigh North Carolina USA, April 2010. ACM.

\bibitem{pougue2021debagreement}
John Pougu{\'e}-Biyong, Valentina Semenova, Alexandre Matton, Rachel Han, Aerin Kim, Renaud Lambiotte, and Doyne Farmer.
\newblock Debagreement: A comment-reply dataset for (dis) agreement detection in online debates.
\newblock In {\em Thirty-fifth Conference on Neural Information Processing Systems Datasets and Benchmarks Track (Round 2)}, 2021.

\bibitem{pougue2023learning}
John Pougu{\'e}-Biyong, Akshay Gupta, Aria Haghighi, and Ahmed El-Kishky.
\newblock Learning stance embeddings from signed social graphs.
\newblock In {\em Proceedings of the Sixteenth ACM International Conference on Web Search and Data Mining}, pages 177--185, 2023.

\bibitem{keuchenius2021important}
Anna Keuchenius, Petter T{\"o}rnberg, and Justus Uitermark.
\newblock Why it is important to consider negative ties when studying polarized debates: A signed network analysis of a dutch cultural controversy on twitter.
\newblock {\em PloS one}, 16(8):e0256696, 2021.

\bibitem{harary1953notion}
Frank Harary.
\newblock On the notion of balance of a signed graph.
\newblock {\em Michigan Mathematical Journal}, 2(2):143--146, 1953.

\bibitem{cartwright1956structural}
Dorwin Cartwright and Frank Harary.
\newblock Structural balance: a generalization of heider's theory.
\newblock {\em Psychological review}, 63(5):277, 1956.

\bibitem{heider1946attitudes}
Fritz Heider.
\newblock Attitudes and cognitive organization.
\newblock {\em The Journal of psychology}, 21(1):107--112, 1946.

\bibitem{davis1967clustering}
James~A Davis.
\newblock Clustering and structural balance in graphs.
\newblock {\em Human relations}, 20(2):181--187, 1967.

\bibitem{aref2019balance}
Samin Aref and Mark~C Wilson.
\newblock Balance and frustration in signed networks.
\newblock {\em Journal of Complex Networks}, 7(2):163--189, 2019.

\bibitem{doreian2009partitioning}
Patrick Doreian and Andrej Mrvar.
\newblock Partitioning signed social networks.
\newblock {\em Social Networks}, 31(1):1--11, 2009.

\bibitem{traag2009community}
Vincent~A Traag and Jeroen Bruggeman.
\newblock Community detection in networks with positive and negative links.
\newblock {\em Physical Review E}, 80(3):036115, 2009.

\bibitem{Kirkley_2019}
Alec Kirkley, George~T. Cantwell, and M.~E.~J. Newman.
\newblock Balance in signed networks.
\newblock {\em Physical Review E}, 99(1), January 2019.

\bibitem{huang2021pole}
Zexi Huang, Arlei Silva, and Ambuj Singh.
\newblock Pole: Polarized embedding for signed networks.
\newblock {\em arXiv preprint arXiv:2110.09899}, 2021.

\bibitem{arefIdentifyingHiddenCoalitions2021}
Samin Aref and Zachary~P. Neal.
\newblock Identifying hidden coalitions in the {{US House}} of {{Representatives}} by optimally partitioning signed networks based on generalized balance.
\newblock 11(1):19939.

\bibitem{babul2024sheep}
Shazia’Ayn Babul and Renaud Lambiotte.
\newblock Sheep, a signed hamiltonian eigenvector embedding for proximity.
\newblock {\em Communications Physics}, 7(1):8, 2024.

\bibitem{arefDetectingCoalitionsOptimally2020}
Samin Aref and Zachary Neal.
\newblock Detecting coalitions by optimally partitioning signed networks of political collaboration.
\newblock 10(1):1506.

\bibitem{eexploratory}
Jan Alyne~Barbosa e~Silva.
\newblock An exploratory study about issue and attribute salience within the system of news promotion, men{\'e}ame.
\newblock 2008.

\bibitem{kaltenbrunner2011comparative}
Andreas Kaltenbrunner, Gustavo Gonzalez, Ricard Ruiz De~Querol, and Yana Volkovich.
\newblock Comparative analysis of articulated and behavioural social networks in a social news sharing website.
\newblock {\em New Review of Hypermedia and Multimedia}, 17(3):243--266, 2011.

\bibitem{gomezLikelihoodbasedFrameworkAnalysis2013}
Vicen{\c c} G{\'o}mez, Hilbert~J. Kappen, Nelly Litvak, and Andreas Kaltenbrunner.
\newblock A likelihood-based framework for the analysis of discussion threads.
\newblock {\em World Wide Web}, 16(5):645--675, November 2013.

\bibitem{aragon2017thread}
Pablo Arag{\'o}n, Vicen{\c{c}} G{\'o}mez, and Andreaks Kaltenbrunner.
\newblock To thread or not to thread: The impact of conversation threading on online discussion.
\newblock In {\em Proceedings of the International AAAI Conference on Web and social media}, volume~11, pages 12--21, 2017.

\bibitem{esmailianCommunityDetectionSigned2015a}
Pouya Esmailian and Mahdi Jalili.
\newblock Community {{Detection}} in {{Signed Networks}}: The {{Role}} of {{Negative}} ties in {{Different Scales}}.
\newblock {\em Scientific Reports}, 5(1):14339, September 2015.

\bibitem{talagaPolarizationMultiscaleStructural2023b}
Szymon Talaga, Massimo Stella, Trevor~James Swanson, and Andreia~Sofia Teixeira.
\newblock Polarization and multiscale structural balance in signed networks.
\newblock {\em Communications Physics}, 6(1):1--15, December 2023.

\bibitem{greenacre2017correspondence}
Michael Greenacre.
\newblock {\em Correspondence analysis in practice}.
\newblock chapman and hall/crc, 2017.

\bibitem{Peralta2024}
Antonio~F. Peralta, Pedro Ramaciotti, J\'anos Kert\'esz, and Gerardo I\~niguez.
\newblock Multidimensional political polarization in online social networks.
\newblock {\em Phys. Rev. Res.}, 6:013170, Feb 2024.

\bibitem{bleiTopicModels2009}
David~M. Blei and John~D. Lafferty.
\newblock Topic {{Models}}.
\newblock In {\em Text {{Mining}}}. {Chapman and Hall/CRC}.

\bibitem{grootendorst2022bertopic}
Maarten Grootendorst.
\newblock Bertopic: Neural topic modeling with a class-based tf-idf procedure.
\newblock {\em arXiv preprint arXiv:2203.05794}, 2022.

\bibitem{gerlachNetworkApproachTopic2018}
Martin Gerlach, Tiago~P. Peixoto, and Eduardo~G. Altmann.
\newblock A network approach to topic models.
\newblock 4(7):eaaq1360.

\bibitem{devlinBERTPretrainingDeep2019}
Jacob Devlin, Ming-Wei Chang, Kenton Lee, and Kristina Toutanova.
\newblock {{BERT}}: {{Pre-training}} of {{Deep Bidirectional Transformers}} for {{Language Understanding}}, May 2019.

\bibitem{hSBM2014}
Tiago~P. Peixoto.
\newblock Hierarchical block structures and high-resolution model selection in large networks.
\newblock {\em Phys. Rev. X}, 4:011047, March 2014.

\bibitem{shiDynamicsSignedNetworks2019}
Guodong Shi, Claudio Altafini, and John~S. Baras.
\newblock Dynamics over {{Signed Networks}}.
\newblock {\em SIAM Review}, 61(2):229--257, January 2019.

\bibitem{Hirschfeld_1935}
H.~O. Hirschfeld.
\newblock A connection between correlation and contingency.
\newblock {\em Mathematical Proceedings of the Cambridge Philosophical Society}, 31(4):520–524, 1935.

\bibitem{benzécri1973analyse}
J.P. Benz{\'e}cri and L.~Bellier.
\newblock {\em L'analyse des donn{\'e}es: Benz{\'e}cri, J.-P. et al. L'analyse des correspondances}.
\newblock L'analyse des donn{\'e}es: le{\c{c}}ons sur l'analyse factorielle et la reconnaissance des formes, et travaux du Laboratoire de statistique de l'Universit{\'e} de Paris VI. Dunod, 1973.

\bibitem{greenacre1984theory}
M.J. Greenacre.
\newblock {\em Theory and Applications of Correspondence Analysis}.
\newblock Academic Press, 1984.

\bibitem{Halford_Prince}
Max Halford.
\newblock {Prince}.

\bibitem{harmonTweepy2023}
Harmon, Joshua Roesslein, and {Other Contributors}.
\newblock Tweepy.
\newblock Zenodo, April 2023.

\bibitem{fruchtermanGraphDrawingForcedirected1991}
Thomas M.~J. Fruchterman and Edward~M. Reingold.
\newblock Graph drawing by force-directed placement.
\newblock {\em Software: Practice and Experience}, 21(11):1129--1164, 1991.

\bibitem{reimersSentenceBERTSentenceEmbeddings2019}
Nils Reimers and Iryna Gurevych.
\newblock Sentence-{{BERT}}: {{Sentence Embeddings}} using {{Siamese BERT-Networks}}.

\bibitem{Somosnlphackathon2022ParaphrasespanishdistilrobertaHugging2022}
Somosnlp-hackathon-2022/paraphrase-spanish-distilroberta · {{Hugging Face}}.

\bibitem{2018arXivUMAP}
L.~{McInnes}, J.~{Healy}, and J.~{Melville}.
\newblock {UMAP: Uniform Manifold Approximation and Projection for Dimension Reduction}.
\newblock {\em ArXiv e-prints}, February 2018.

\bibitem{McInnes2017hdbscan}
Leland McInnes, John Healy, and Steve Astels.
\newblock hdbscan: Hierarchical density based clustering.
\newblock {\em The Journal of Open Source Software}, 2(11), March 2017.

\bibitem{scikit-learn}
F.~Pedregosa, G.~Varoquaux, A.~Gramfort, V.~Michel, B.~Thirion, O.~Grisel, M.~Blondel, P.~Prettenhofer, R.~Weiss, V.~Dubourg, J.~Vanderplas, A.~Passos, D.~Cournapeau, M.~Brucher, M.~Perrot, and E.~Duchesnay.
\newblock Scikit-learn: {{Machine}} learning in {{Python}}.
\newblock {\em Journal of Machine Learning Research}, 12:2825--2830, 2011.

\bibitem{bird2009natural}
Steven Bird, Edward Loper, and Ewan Klein.
\newblock {\em Natural Language Processing with Python}.
\newblock O'Reilly Media Inc., 2009.

\bibitem{Honnibal_spaCy_Industrial-strength_Natural_2020}
Matthew Honnibal, Ines Montani, Sofie Van~Landeghem, and Adriane Boyd.
\newblock {spaCy: Industrial-strength Natural Language Processing in Python}.
\newblock 2020.

\bibitem{calinski1974dendrite}
Tadeusz Cali{\'n}ski and Jerzy Harabasz.
\newblock A dendrite method for cluster analysis.
\newblock {\em Communications in Statistics-theory and Methods}, 3(1):1--27, 1974.

\end{thebibliography}

\clearpage
\appendix

\renewcommand\thefigure{A\arabic{figure}}  
\renewcommand\thetable{A\arabic{table}}  
\section{Appendix}\label{appendix}
\subsection{Reciprocal voting analysis}
We analyzed mutual voting behavior between users for the topics Russia and Politics. For each user pair, we calculated the number and mean of votes in each direction, retaining only pairs with at least three votes both ways. Tables~\ref{tab:crosstab_russia}-~\ref{tab:crosstab_politics} show a strong degree of agreement and symmetry between users' voting patterns.

\begin{table}[h]
\centering
\caption{Cross-tabulation of binned average vote scores (Topic: \textit{Russia})}
\begin{tabular}{lccccc}
\toprule
 & (-1.01, -0.899] & (-0.899, -0.3] & (-0.3, 0.3] & (0.3, 0.899] & (0.899, 1.01] \\
\midrule
(-1.01, -0.899] & 116 & 2 & 2 & 1 & 0 \\
(-0.899, -0.3]  & 2   & 2 & 1 & 0 & 1 \\
(-0.3, 0.3]     & 2   & 1 & 0 & 1 & 1 \\
(0.3, 0.899]    & 1   & 0 & 1 & 0 & 17 \\
(0.899, 1.01]   & 0   & 1 & 1 & 17 & 2892 \\
\bottomrule
\end{tabular}
\label{tab:crosstab_russia}
\end{table}

\begin{table}[h]
\centering
\caption{Cross-tabulation of binned average vote scores (Topic: \textit{BroadPolitics})}
\begin{tabular}{lccccc}
\toprule
 & (-1.01, -0.899] & (-0.899, -0.3] & (-0.3, 0.3] & (0.3, 0.899] & (0.899, 1.01] \\
\midrule
(-1.01, -0.899] & 96 & 15 & 6 & 4 & 3 \\
(-0.899, -0.3]  & 15 & 8  & 4 & 5 & 6 \\
(-0.3, 0.3]     & 6  & 4  & 0 & 5 & 5 \\
(0.3, 0.899]    & 4  & 5  & 5 & 24 & 96 \\
(0.899, 1.01]   & 3  & 6  & 5 & 96 & 5060 \\
\bottomrule
\end{tabular}
\label{tab:crosstab_politics}
\end{table}

\clearpage

\subsection{BERTopic algorithm}\label{appendix:bertopic}
In this section, we describe the steps followed by the BERTopic algorithm and the rationale behind each parameter choice:
\begin{enumerate} 
    \item \textbf{Sentence embeddings}: This step transforms a text corpus into a collection of vectors, where each vector identifies a text in a multidimensional space. We use SentenceTransformers (\cite{reimersSentenceBERTSentenceEmbeddings2019}), a variation of the BERT (Bidirectional Encoder Representations from Transformers) architecture specialized for creating sentence embeddings. In particular, given that our text corpus is in Spanish, we choose a model \textit{paraphrase-spanish-distilroberta} (\cite{Somosnlphackathon2022ParaphrasespanishdistilrobertaHugging2022}) pre-trained on Spanish text data. The outputs are 768-dimensional vectors for each document.
    
    \item \textbf{Dimensionality reduction}: Given the high dimensionality of the vectors produced by the previous step, we need to represent the corpus in a lower-dimensional space before proceeding with the clustering task. We use the default technique of BERTopic, UMAP (\cite{2018arXivUMAP}) (Uniform Manifold Approximation and Projection), choosing the parameters such as \textit{$\text{n\_neighbors} = 50$}, which is a parameter that aims to balance the importance of local (low values) versus global (high values) patterns in the data; \textit{$\text{n\_components} = 75$}, which is a parameter that controls the dimension of the output vectors, to reduce to approximately $10 \%$ of the initial dimensions; and last, we chose as similarity metrics the cosine similarity\footnote{Given two vectors $\vec{a}$ and $\vec{b}$, the cosine similarity is defined as $\frac{\vec{a} \cdot \vec{b}}{||\vec{a}|| ||\vec{b}||}$.}.
    
    \item \textbf{Clustering}: after reducing the text embeddings to a lower dimensionality with UMAP, we cluster them using HDBSCAN (\cite{McInnes2017hdbscan}) (Hierarchical Density-Based Spatial Clustering of Applications with Noise) algorithm. We chose the cosine similarity as metrics again, then we set \textit{$\text{min\_cluster\_size} = 50$}, to avoid the presence of too granular topics (i.e., few stories assigned per topic). Two parameters critical in determining cluster sizes and number of outliers are $\textit{min\_sample}$ and $\textit{cluster\_selection\_epsilon}$. $\textit{min\_sample}$ determines how conservative a clustering procedure should be, i.e., if high, more texts will be found as outliers; $\textit{cluster\_selection\_epsilon}$ defines the radius within which two clusters will be merged. \ref{appendix:sensitivity-analysis} shows the performed tests to understand how the results are affected by this parameter choice. We finally chose \textit{$\text{min\_sample}=1$} and \textit{$\text{cluster\_selection\_epsilon}=3 \times 10^{-6}$}, which ensures a trade-off between the number of topics and outliers.

    \item \textbf{Vectorizer}: In the previous steps, we subdivided the text corpus into groups. From now on, we want to represent each topic with relevant keywords. Here, we use CountVectorizer from the \textit{sci-kit learn} package (\cite{scikit-learn}) to convert the text corpus of each topic into a matrix of token counts (i.e., we want to find the most popular words for each topic). We remove the Spanish stopwords, we limit the number of features to 1000, and we consider words appearing more than 100 times.

    \item \textbf{c-TFIDF}: from the previous step, we obtained matrices of frequencies of each word in each document of a certain topic. Here, we want to find the most relevant words per topic; therefore, we use the c-TFIDF, a modified version of TFIDF that accounts for a topic-level measurement instead of a document-level measurement.

    \item \textbf{Representation Tuning}: given the keywords for each topic, we want to perform an additional step to bind the results steps (1)--(3) to the ones of steps (4)--(5). Until now, the same keywords could be the most relevant for all the topics, as we determined them for each topic independently. We employ the method \textit{MaximalMarginalRelevance} to maximize the diversity of those keywords.

\end{enumerate} 
\subsubsection{Sensitivity analysis of HDBSCAN parameters}\label{appendix:sensitivity-analysis}
\begin{figure}
    \centering
    \includegraphics[width=\textwidth]{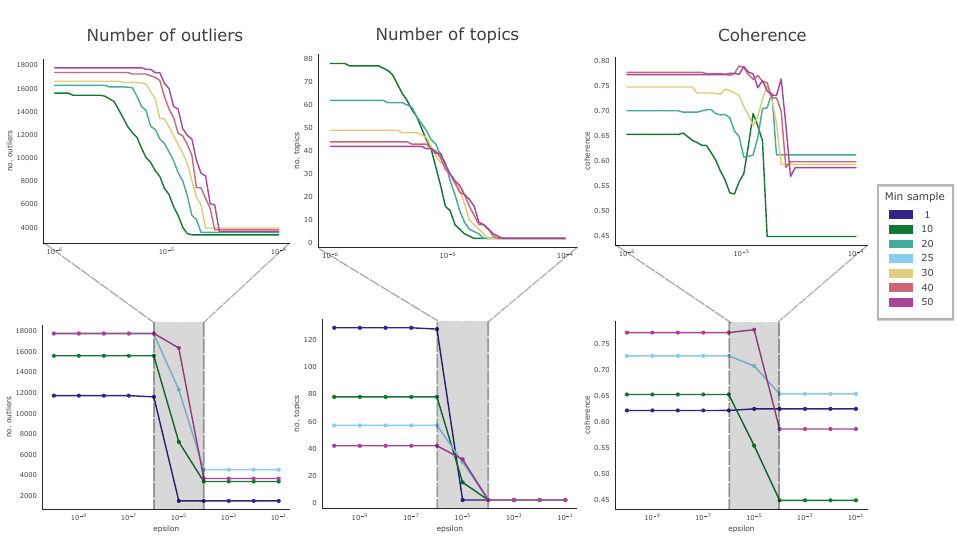}
    \caption{Variation of the number of outliers, the number of topics, and the coherence, respectively varying the parameters $\epsilon$ and min sample. As we would like a situation with not many outliers and a 'reasonable' number of topics, we choose the following values for the parameters: \textit{$\text{min\_sample}=1$} and \textit{$\text{cluster\_selection\_epsilon}=3 \times 10^{-6}$}.}
    \label{fig:sensitivity-analysis}
\end{figure}

Given the high number of model parameters that need fine-tuning, we performed a sensitivity analysis for the clustering step of the BERTopic algorithm. The clustering is performed using the HDBSCAN method, which has, among others, two parameters: $\textit{cluster\_selection\_epsilon}$ and $\textit{min\_sample}$. Figure~\ref{fig:sensitivity-analysis} shows the results for the number of outliers, topics, and the coherence value, ranging for several values of $\textit{cluster\_selection\_epsilon}$ and $\textit{min\_sample}$.

\clearpage
\subsection{hSBM Topic Model steps}\label{appendix:hsbm}
In this section, we describe the steps followed to perform the hSBM topic modeling:
\begin{enumerate}
    \item \textbf{Text preprocessing}: We remove the stopwords using the NLTK (\cite{bird2009natural}) Spanish stopwords list and create tokens using the Spacy package (\cite{Honnibal_spaCy_Industrial-strength_Natural_2020}).
    \item \textbf{Graph creation}: We create the word-document bipartite graph, where the edge weights are given by the frequency of each word in each document.
    \item \textbf{hSBM fit}: We fit the community detection model on the word-document network, resulting in hierarchical subdivisions of words and documents into non-overlapping clusters.
    \item \textbf{Topic representation}: Similarly to BERTopic, we apply the CountVectorizer and compute the c-TF-IDF. Then, we identify each topic's top 10 highest-scoring terms and assign them as topic representations.
\end{enumerate}

\clearpage

\subsection{Topic Modelling results}
In this section, we show the results of the intermediate topic modeling, before fine-tuning.
Tables~\ref{tab:topics-bertopic} and~\ref{tab:topics-hsbm} show the topic representations obtained with BERTopic and TM-hSBM, respectively.
\begin{longtblr}[
  label = {tab:topics-bertopic},
  entry = none,
  caption = {topics obtained with bertopic before any outlier reduction},
]{
  hline{1-2,184} = {-}{},
}
\textbf{Topic} & \textbf{No. of stories} & \textbf{Topic Representation}\\
-1 & 16633 & juan, empresa, ayuntamiento, china \\
0 & 2734 & ucrania, rusia, ruso, ucraniano \\
1 & 2129 & elecciones, voto, votos, electoral \\
2 & 1754 & perros, animales, animal, arqueologos \\
3 & 1573 & novela, videojuegos, videojuego, libros \\
4 & 1284 & policia, agentes, arrestado, agresion \\
5 & 1085 & prision, condena, condenado, delito \\
6 & 1003 & musica, cancion, banda, musical \\
7 & 979 & bancos, inflacion, interes, deuda \\
8 & 882 & huelga, medicos, hospitales, urgencias \\
9 & 769 & temperaturas, temperatura, climatico, calentamiento \\
10 & 754 & luna, mision, universo, planeta \\
11 & 752 & futbol, jugadores, deporte, liga \\
12 & 700 & electricidad, renovables, hidrogeno, solar \\
13 & 619 & laboral, trabajadores, trabajador, empleados \\
14 & 572 & chatgpt, inteligencia, gpt, lenguaje \\
15 & 533 & alquiler, viviendas, pisos, propietarios \\
16 & 520 & miguel, antonio, juan, alberto \\
17 & 503 & coches, vehiculos, automoviles, coche \\
18 & 488 & violencia, prostitucion, feminista, sexuales \\
19 & 418 & carne, comida, alimentos, productos \\
20 & 403 & aborto, embarazo, castilla, protocolo \\
21 & 362 & donana, aguas, ecologica, cuenca \\
22 & 361 & alumnos, educacion, profesores, escuelas \\
23 & 356 & musk, tuit, mensajes, publicidad \\
24 & 348 & fotografia, artista, fotografias, fotos \\
25 & 341 & israel, israeli, civiles, soldados \\
26 & 327 & constitucional, cgpj, reforma, tribunal \\
27 & 291 & coronavirus, muertes, enfermedad, pandemia \\
28 & 288 & cocaina, medicamentos, droga, drogas \\
29 & 285 & whatsapp, proteccion, microsoft, informacion \\
30 & 282 & periodismo, desinformacion, profesion, news \\
31 & 281 & manifestantes, protesta, activistas, protestas \\
32 & 279 & puerto, barcos, aguas, costa \\
33 & 252 & impuesto, impuestos, fiscal, tributaria \\
34 & 241 & vehiculo, accidente, coche, trafico \\
35 & 239 & transportes, transporte, pasajeros, movilidad \\
36 & 230 & cancer, enfermedad, enfermedades, cientificos \\
37 & 229 & incendios, incendio, bomberos, fuego \\
38 & 228 & jubilacion, reforma, ipc, ingresos \\
39 & 225 & lengua, castellano, palabra, catalan \\
40 & 199 & fallecio, murio, actriz, cancer \\
41 & 197 & amazon, moviles, telefonica, microsoft \\
42 & 197 & franquista, victimas, democratica, historica \\
43 & 194 & iglesia, abuso, victimas, sexualmente \\
44 & 186 & imperio, batalla, reina, soldados \\
45 & 169 & comision, donana, censura, calle \\
46 & 168 & sexual, sexuales, relaciones, mujeres \\
47 & 168 & vuelos, aviones, piloto, aeropuerto \\
48 & 165 & sueldo, concejales, alcalde, salario \\
49 & 165 & economicos, supermercados, inflacion, economica \\
50 & 154 & fiscalia, corrupcion, investiga, fraude \\
51 & 154 & instagram, perfil, contenido, internet \\
52 & 147 & migrantes, inmigrantes, frontera, mexico \\
53 & 143 & hospital, urgencias, medico, paciente \\
54 & 140 & supermercados, alimentos, precios, compra \\
55 & 140 & terremoto, turquia, siria, costa \\
56 & 140 & criptomonedas, inversores, digitales, dolares \\
57 & 136 & bancaria, clientes, empleo, instancia \\
58 & 134 & palacio, presupuesto, ultraderechista, rodriguez \\
59 & 133 & enfermedades, medicamentos, sangre, enfermedad \\
60 & 133 & gonzalez, polonia, espionaje, vicepresidente \\
61 & 131 & empleados, recortes, trabajadores, microsoft \\
62 & 130 & instagram, comida, lugares, contenidos \\
63 & 130 & bebe, hermano, foto, propias \\
64 & 129 & nazis, alemanes, detenidos, soldados \\
65 & 129 & espectadores, television, tve, tv \\
66 & 128 & suicidio, adolescentes, muertes, causa \\
67 & 127 & iglesia, articulos, publicar, cientificos \\
68 & 127 & china, oeste, central, provincia \\
69 & 125 & pobreza, poblacion, porcentaje, estadistica \\
70 & 121 & marruecos, occidental, autonomia, relaciones \\
71 & 121 & iglesia, catolica, basura, ferrovial \\
72 & 118 & venezuela, reuniones, colombia, xunta \\
73 & 116 & franquismo, infantil, trabajar, ultraderecha \\
74 & 112 & chino, aerea, espionaje, misiles \\
75 & 111 & digital, moviles, instagram, tecnologias \\
76 & 110 & arboles, especies, verdes, reserva \\
77 & 108 & licencia, edificios, ayuntamiento, viviendas \\
78 & 104 & censura, candidato, sanchez, debate \\
79 & 103 & fiestas, festival, navidad, sanidad \\
80 & 103 & arabia, saudi, siria, relaciones \\
81 & 102 & franco, flores, franquista, negociacion \\
82 & 102 & trump, expresidente, jurado, republicano \\
83 & 102 & bolsonaro, lula, brasil, expresidente \\
84 & 101 & miedo, relaciones, bienestar, individuo \\
85 & 99 & periodismo, droga, sovietica, incendios \\
86 & 99 & japon, literatura, historias, dirigida \\
87 & 96 & residuos, contaminacion, toneladas, ambiental \\
88 & 96 & memoria, financiera, riqueza, dispositivos \\
89 & 95 & acoso, profesor, escuela, alumnos \\
90 & 94 & netflix, tiktok, anuncios, contenido \\
91 & 93 & cientifico, cientifica, representacion, dispositivos \\
92 & 93 & hogares, familias, pobreza, ingresos \\
93 & 93 & hollywood, humanos, derechos, agente \\
94 & 91 & peru, protestas, virgen, abogados \\
95 & 90 & riqueza, empresarios, economica, empresarial \\
96 & 89 & trafico, vehiculo, coche, sancion \\
97 & 89 & contratos, millon, euros, contrato \\
98 & 87 & entrevista, escritor, sanchez, habla \\
99 & 86 & revista, publicacion, felipe, asociacion \\
100 & 86 & suiza, franquismo, prostitucion, investigado \\
101 & 86 & contaminacion, sustancias, salud, muertes \\
102 & 86 & concierto, festival, responsabilidad, amenazas \\
103 & 85 & motor, motores, combustible, pobreza \\
104 & 85 & mapa, urbano, corea, empresarios \\
105 & 84 & comisaria, martinez, acusa, sancion \\
106 & 84 & turquia, comicios, elecciones, independencia \\
107 & 83 & lugares, hijos, familiares, comentarios \\
108 & 83 & europea, bruselas, conservadores, politicos \\
109 & 81 & dolares, nacion, envio, suiza \\
110 & 81 & viral, javier, radio, cadiz \\
111 & 81 & ruta, sierra, montana, naturales \\
112 & 81 & muerte, condenado, protestas, ejecucion \\
113 & 79 & parlamento, europeo, marruecos, escandalo \\
114 & 79 & microsoft, software, lenguaje, sistemas \\
115 & 79 & enfermedades, especies, enfermedad, humanos \\
116 & 79 & padre, murio, felipe, franquismo \\
117 & 78 & restaurante, carbon, peru, barrio \\
118 & 78 & bruselas, educacion, odio, restricciones \\
119 & 77 & imagen, ciudadano, alfonso, diversos \\
120 & 76 & republicanos, republicano, news, expresidente \\
121 & 75 & conflictos, invasion, civiles, george \\
122 & 75 & ilegal, detencion, pablo, inmigrantes \\
123 & 74 & agricultura, represion, contratos, socialistas \\
124 & 74 & supermercados, produccion, agricultura, vender \\
125 & 74 & colores, lluvias, simbolo, palacio \\
126 & 74 & italia, republica, dictadura, franco \\
127 & 74 & ejercicio, actividad, adolescentes, bienestar \\
128 & 74 & radio, contenidos, rtve, documental \\
129 & 72 & espionaje, republicano, normativa, conservadores \\
130 & 71 & ortega, nino, lengua, bolsonaro \\
131 & 71 & reina, palacio, felipe, cruz \\
132 & 70 & racismo, racistas, discriminacion, nazis \\
133 & 70 & mexico, lopez, venezuela, chile \\
134 & 70 & valladolid, marruecos, cancer, suicidio \\
135 & 68 & negocios, trabajar, sector, condiciones \\
136 & 68 & news, francesa, fraude, argentina \\
137 & 67 & peru, bolsonaro, brasil, dictadura \\
138 & 65 & tareas, dictadura, imperio, brasil \\
139 & 65 & economicas, clases, revolucion, cultural \\
140 & 65 & drones, costa, policias, asociaciones \\
141 & 64 & reaccion, comentarios, discriminacion, ultraderechista \\
142 & 64 & artistas, eta, viva, academia \\
143 & 64 & reto, daniel, positivo, cristina \\
144 & 64 & odio, tecnica, acoso, ultraderechista \\
145 & 64 & tratado, donana, representan, multinacional \\
146 & 63 & sueldo, edificios, cronica, juventud \\
147 & 63 & tecnico, gastos, barcelona, financiero \\
148 & 63 & historicos, restaurante, franco, novela \\
149 & 63 & alemanes, america, extremadura, hipotecas \\
150 & 62 & arquitectura, carbono, economicas, cientifico \\
151 & 60 & moral, policial, protestas, religion \\
152 & 60 & arboles, parque, ampliacion, obras \\
153 & 59 & pensamiento, conocimiento, clasico, duda \\
154 & 59 & bildu, terrorismo, daba, palabras \\
155 & 59 & movilidad, trafico, ciudades, automoviles \\
156 & 59 & nazi, qatar, oferta, britanica \\
157 & 59 & arquitectura, edificios, torres, construccion \\
158 & 58 & norma, sexual, tribunales, vigor \\
159 & 58 & corto, demasiado, vineta, minimo \\
160 & 58 & competencia, telefonica, consumidores, comision \\
161 & 57 & electoral, martinez, junta, fernandez \\
162 & 57 & trafico, limites, coches, accidente \\
163 & 57 & libros, digital, dificultades, pensar \\
164 & 57 & inversores, inversion, prevencion, efectivo \\
165 & 57 & consumidores, ciento, hielo, comercial \\
166 & 56 & comision, bruselas, europea, deuda \\
167 & 56 & nazis, extrema, alemanes, odio \\
168 & 55 & canciones, sindicato, laborales, municipal \\
169 & 55 & odio, virgen, agresiones, expresion \\
170 & 55 & armada, saudi, construccion, cristina \\
171 & 54 & parlamento, terroristas, pobres, relaciones \\
172 & 54 & coalicion, republica, presidencia, franquismo \\
173 & 54 & indemnizacion, videojuegos, galicia, condena \\
174 & 54 & salamanca, colombia, regional, represion \\
175 & 53 & conservador, financiacion, polonia, territorios \\
176 & 53 & salarios, salario, inflacion, sueldo \\
177 & 53 & prostitucion, republicanos, electoral, alcaldia \\
178 & 52 & victimas, terrorismo, bildu, tratamiento \\
179 & 52 & michael, reyes, jugador, teoria \\
180 & 51 & cerebro, memoria, laboratorio, conciencia
\end{longtblr}

\begin{longtblr}[
  label = {tab:topics-hsbm},
  entry = none,
  caption ={topics obtained with hsbm}
]{
  hline{1-2,141} = {-}{},
}
\textbf{Level} & \textbf{Topic} & \textbf{Representation}\\
0 & 0 & energetico, eolico, gw, electricidad, fotovoltaico\\
0 & 1 & vladimir, ucraniano, paz, invasion, ruso\\
0 & 2 & recep, arabia, magnitud, erdogar, siria\\
0 & 3 & cinematografico, guionista, serie, ficcion, hollywood\\
0 & 4 & streaming, telefono, meta, web, internet\\
0 & 5 & alec, yeremi, biyin, fast, payaso\\
0 & 6 & faso, irak, velo, arma, mahsa\\
0 & 7 & complutense, curso, facultad, estudiante, educacion\\
0 & 8 & fascismo, adolf, republica, holocausto, fascista\\
0 & 9 & lago, metro, kilometro, oceano, marino\\
0 & 10 & comisario, mediador, juez, audiencia, presunto\\
0 & 11 & escritor, calle, aquel, san, nombre\\
0 & 12 & operacion, detener, hachis, civil, kilo\\
0 & 13 & salud, centro, atender, clinico, hospitalario\\
0 & 14 & pandemia, enfermedad, depresion, anar, poblacion\\
0 & 15 & precio, alquilar, inmueble, inquilino, propietario\\
0 & 16 & generativo, herramienta, chatbot, lenguaje, gpt\\
0 & 17 & mohamed, saharauis, polisario, eurodiputado, rabat\\
0 & 18 & partido, candidato, pp, elección, ayuntamiento\\
0 & 19 & comunidad, facultativo, pediatra, consejeria, huelga\\
0 & 20 & sabor, menu, comer, alimento, patata\\
0 & 21 & senor, gustar, alguien, nadie, querer\\
0 & 22 & deportivo, jugar, copa, futbolista, torneo\\
0 & 23 & luis, baltar, jo, ourense, consistorio\\
0 & 24 & television, tve, radio, tv, mediaset\\
0 & 25 & inversor, bitcoin, bancario, entidad, suisse\\
0 & 26 & demasiado, tira, corto, caracter, manel\\
0 & 27 & habiar, familia, escuela, bebe, nino\\
0 & 28 & hablar, siempre, alguien, aprender, gente\\
0 & 29 & vox, discurso, democracia, partido, politica\\
0 & 30 & album, canción, concierto, cantante, musical\\
0 & 31 & ley, embarazo, aborto, derecho, tran\\
0 & 32 & bajmut, bakhmut, rusia, soldado, donetsk\\
0 & 33 & natalidad, elevado, rico, riqueza, economia\\
0 & 34 & hallazgo, ac, bronce, antiguedad, piedra\\
0 & 35 & tierra, foto, cielo, elemento, paisaje\\
0 & 36 & it, you, for, cancion, on\\
0 & 37 & autor, literario, historia, artista, escritor\\
0 & 38 & sequia, hielo, cambio, calor, oceano\\
0 & 39 & satelit, spacex, nave, utc, luna\\
0 & 40 & condenar, victima, abuso, agresion, delito\\
0 & 41 & rusia, petrolero, licuado, exportación, gnl\\
0 & 42 & trabajador, paro, cgt, sindical, patronal\\
0 & 43 & abc, variación, error, exagerar, economista\\
0 & 44 & oiea, planta, residuo, fukushima, energia\\
0 & 45 & gaza, judio, benjamin, palestina, cisjordania\\
0 & 46 & pozo, andalucia, hidrico, parque, rio\\
0 & 47 & patrimonio, renta, tributo, fortuna, sucesión\\
0 & 48 & crucero, titanic, velero, embarcacion, orca\\
0 & 49 & mosquito, microorganismo, genoma, especie, humano\\
0 & 50 & kmh, circular, coche, radar, carretera\\
0 & 51 & calle, detener, local, incidente, disparo\\
0 & 52 & marca, bmw, combustion, tesla, fabricante\\
0 & 53 & salario, ingreso, jornada, empleado, despido\\
0 & 54 & aplicacion, ley, penal, condenado, rebajar\\
0 & 55 & formigal, pirineo, ecologicar, sostenible, ribera\\
0 & 56 & barrio, edificio, calle, habitante, vecino\\
0 & 57 & consello, monbus, contrato, xunto, tsxg\\
0 & 58 & derribar, espia, pentagono, tripulado, dron\\
0 & 59 & filtracion, wikileaks, sabotaje, assange, cia\\
0 & 60 & internet, periodismo, digital, idea, social\\
0 & 61 & acción, anterior, recaudar, cifra, neto\\
0 & 62 & contagio, granja, hn, oms, coronavirus\\
0 & 63 & episcopal, santo, vaticano, diocesis, papa\\
0 & 64 & mexicano, venezolano, guaido, mexico, obrador\\
0 & 65 & texas, aborto, disney, democrata, legislador\\
0 & 66 & euskadi, oskar, otegi, abertzale, navarra\\
0 & 67 & joe, capitolio, carlson, news, republicano\\
0 & 68 & macarén, iglesia, abascal, pablo, ramon\\
0 & 69 & partido, popular, lider, investidura, pedro\\
0 & 70 & plataforma, mastodon, red, blue, tuit\\
0 & 71 & atari, rol, spectrum, ordenador, nintendo\\
0 & 72 & vii, muralla, antiguo, ac, dc\\
0 & 73 & trastorno, cerebro, medicamento, vacuna, farmaco\\
0 & 74 & insecto, lobo, iberico, fauna, conservacion\\
0 & 75 & reforma, judicial, organo, vocal, tribunal\\
0 & 76 & desquadra, detenido, agredir, hombre, agresion\\
0 & 77 & recuperacion, der, von, comunitario, brusela\\
0 & 78 & eeuu, corea, bric, xi, jinping\\
0 & 79 & sargento, acuartelamiento, tejera, jarava, cuartel\\
0 & 80 & sector, empresa, ferrovial, mercado, inversion\\
0 & 81 & tormenta, meteorologia, calido, aemet, lluvia\\
0 & 82 & galardon, gala, sabbath, nobel, nominado\\
0 & 83 & consumidor, compra, producto, mercadona, carrefour\\
0 & 84 & pasajero, ferrocarril, trayecto, transporte, via\\
0 & 85 & ministro, anticrisi, consejo, decreto, aprobar\\
0 & 86 & borbon, monarca, abu, infanta, reina\\
0 & 87 & estadistico, economia, tasa, decima, crecimiento\\
0 & 88 & vox, tuberculosis, bovino, latido, junta\\
0 & 89 & mojacar, comicio, correos, melilla, jec\\
0 & 90 & pacma, cinegetico, bienestar, jabali, animalista\\
0 & 91 & software, linux, aplicación, desarrollador, microsoft\\
0 & 92 & mercado, electricidad, regulado, mayorista, gas\\
0 & 93 & tierra, espacial, particula, webb, agujero\\
0 & 94 & ruanda, boris, johnson, irlanda, rishi\\
0 & 95 & derogar, aprobo, unidas, diputado, aprobar\\
0 & 96 & contenciosoadministrativo, indemnizacion, juzgado, indemnizar, recurso\\
0 & 97 & explosion, transportar, quimico, descarrilar, southern\\
0 & 98 & adquisitivo, discontinuo, bruto, subida, minimo\\
0 & 99 & secretaria, violencia, ley, irenir, feminista\\
0 & 100 & votante, sociologica, sondeo, elección, barometro\\
0 & 101 & ponsati, junquera, independentista, borra, proz\\
0 & 102 & planta, reciclar, co, tonelada, envase\\
0 & 103 & aguirre, pp, yolanda, madrilén, comunidad\\
0 & 104 & arder, roda, evacuar, monte, llama\\
0 & 105 & morada, coalicion, belarra, vicepresidenta, montero\\
0 & 106 & paulo, inacio, luiz, brasileno, brasilia\\
0 & 107 & provincial, acusado, condenar, fiscalia, condena\\
0 & 108 & infanteria, entrenamiento, abrams, ucrania, blindado\\
0 & 109 & deposito, central, inflacion, tipo, euribor\\
0 & 110 & scholz, bukelir, naciones, violación, onu\\
0 & 111 & vladimir, putin, paramilitar, ruso, yevgeny\\
0 & 112 & pienso, leche, cosecha, cereal, cultivo\\
0 & 113 & arbitral, bartomeu, negreirar, rfef, fc\\
1 & 0 & grado, mas, solar, mar, espacial\\
1 & 1 & invasion, otan, militar, putin, trump\\
1 & 2 & civil, peru, presidente, brasil, derechos\\
1 & 3 & tener, vida, hacer, mas, ser\\
1 & 4 & compania, plataforma, millón, electrico, usuario\\
1 & 5 & aviar, emerito, gripe, juan, felipe\\
1 & 6 & alquiler, sanitario, hospital, sindicato, laboral\\
1 & 7 & si, él, derecho, jo, violencia\\
1 & 8 & dos, padre, él, jugador, habiar\\
1 & 9 & menor, victima, guardia, hombre, agente\\
1 & 10 & tumba, yacimiento, imperio, arqueologico, ac\\
1 & 11 & sanchez, ayuso, alberto, nunez, electoral\\
1 & 12 & gas, mwh, ohio, fontdevila, grafico\\
1 & 13 & ave, universo, humano, estrella, telescopio\\
1 & 14 & software, apple, pc, ordenador, windows\\
1 & 15 & economia, trimestre, euros, subida, petroleo\\
1 & 16 & estadounidense, china, wagner, rusia, incendio\\
1 & 17 & regadio, agricultor, parque, alimento, sequia\\
1 & 18 & gobierno, galicia, xunta, fiscal, reforma\\
1 & 19 & bildu, partido, igualdad, congreso, tribunal\\
2 & 0 & primero, si, poder, hacer, tener\\
2 & 1 & hacer, tener, euros, romano, precio\\
2 & 2 & ley, madrid, vox, él, poder\\
2 & 3 & rusia, ucrania, ruso, primero, europeo\\
3 & 0 & primero, si, poder, hacer, tener
\end{longtblr}

\begin{figure}
    \centering
    \includegraphics[width=1\linewidth]{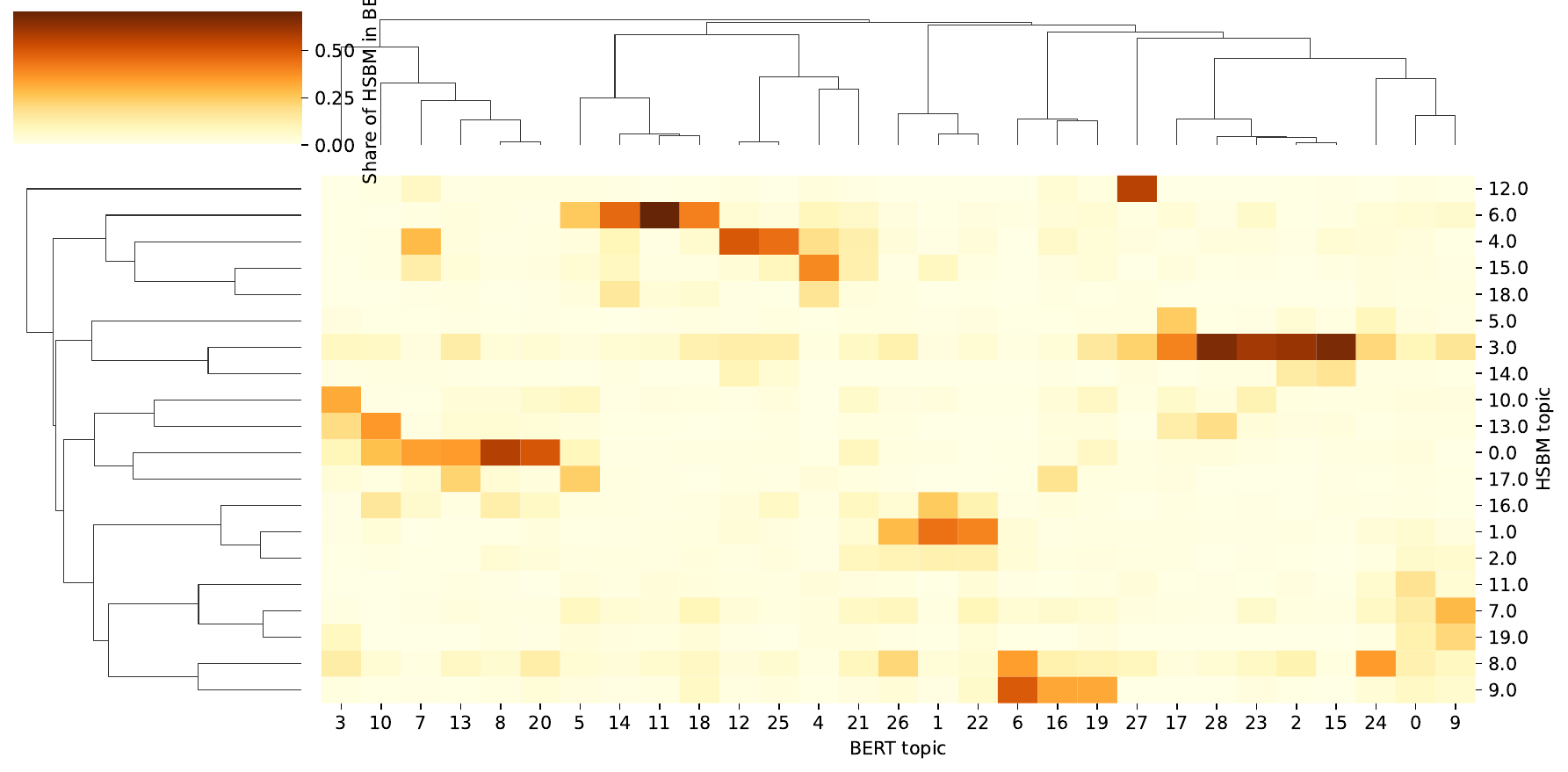}
    \caption{Confusion matrix with colors indicating the overlap between topics identified by BERTopic and hSBM.}
    \label{fig:confusion-mat}
\end{figure}

\clearpage
\subsection{Ideology from Twitter} \label{app:twitter}
The study analyzed tweets from the following political parties and their associated Twitter accounts:
\begin{itemize}
   \item PACMA (@partidopacma) (Animalist Party Against Mistreatment of Animals): @lau\_duart, @sanchezcastejon, @yolanda\_morpe, @crisgarsalazar
    \item Más País (@maspais\_es): @ierrejon, @Monica\_Garcia\_G, @EduardoFRub, @isabanes, @Rita\_Maestre, @Equo, @P\_GomezPerpinya, @compromis, @htejero\_, @MasPais\_Es, @MasMadrid\_\_
    \item VOX (@vox\_es): @Santi\_ABASCAL, @Jorgebuxade, @Ortega\_Smith, @ivanedlm, @monasterioR, @Igarrigavaz, @juan\_ggallardo, @MeerRocio, @VOX\_Congreso, @\_patricia\_rueda
    \item IU (@izquierdaunida): @agarzon, @sirarego, @EnriqueSantiago, @iuandalucia, @InmaNietoC, @joanmena, @iucyl, @Toni\_Valero, @Congreso\_Es, @ma\_bustamante84, @Roser\_Maestro, @iurioja, @elpce
    \item PODEMOS (@PODEMOS): @PabloIglesias, @IreneMontero, @ionebelarra, @isaserras, @Yolanda\_Diaz\_, @PabloEchenique, @PabloFdez, @MayoralRafa, @AleJacintoUrang, @Pam\_Angela\_, @MazelLilith, @SofCastanon, @VickyRosell, @nachoalvarez\_, @juralde, @jessicaalbiach, @m\_tere\_perez, @JA\_DelgadoRamos
    \item Ciudadanos (@ciudadanoscs): @InesArrimadas, @BalEdmundo, @carrizosacarlos, @GuillermoDiazCs, @begonavillacis, @FranciscoIgea, @CiutadansCs, @PatriciaGuaspB, @jordi\_canyas, @MelisaRguezH, @Beatriz\_Pino\_, @Nmartinblanco
    \item PSOE (Spanish Socialist Workers' Party, @PSOE): @sanchezcastejon, @Adrilastra, @salvadorilla, @mjmonteroc, @felipe\_sicilia, @NadiaCalvino, @carmencalvo\_, @abalosmeco, @isabelrguez, @Pilar\_Alegria, @\_JuanEspadas, @luistudanca, @santicl
    \item PP (Partido Popular, @ppopular): @pablocasado\_, @TeoGarciaEgea, @cucagamarra, @NunezFeijoo, @IdiazAyuso, @Aglezterol, @AlmeidaPP\_, @alferma1, @abeltran\_ana, @alejandroTGN, @eliasbendodo, @anapastorjulian, @erodriguez\_2019, @JuanMa\_Moreno, @jaimedeolano
\end{itemize}

\begin{figure}[ht!]
    \centering
    \includegraphics[width=0.8\linewidth]{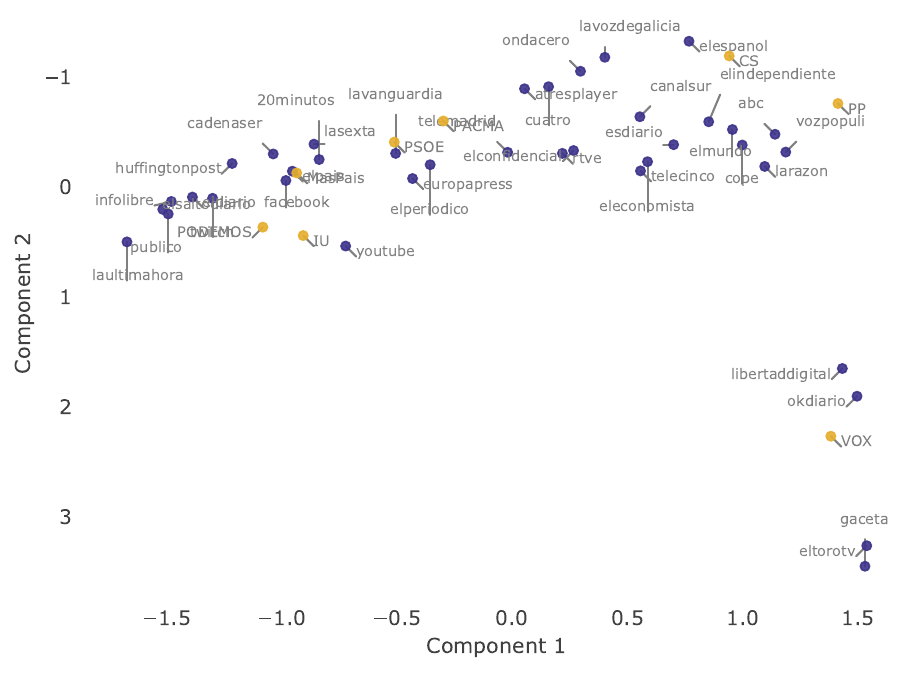}
    \caption{Visualization of the first two components of the correspondence analysis for both news outlets (dark purple) and political parties (yellow). We interpret the first dimension as left-right ideology, and the second dimension (not used in this paper) as mainstream-radical.}
    \label{fig:app:twitter}
\end{figure}

\begin{figure}[ht!]
    \centering
    \includegraphics[width=0.7\linewidth]{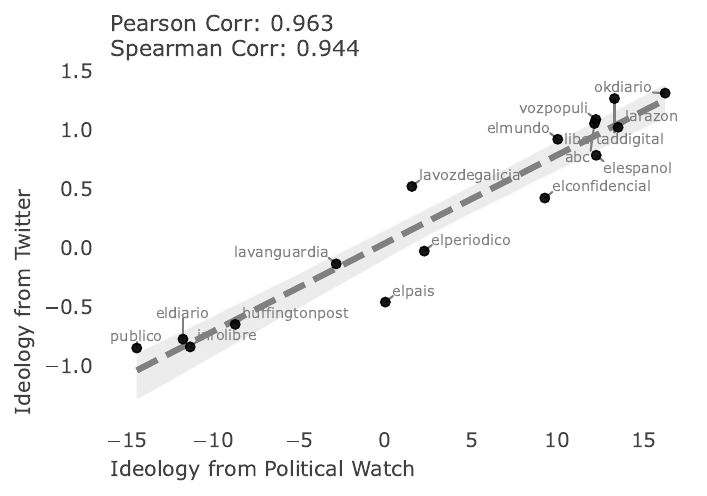}
    \caption{Comparison of the ideology of media outlets and Twitter accounts. The x-axis represents the ideology of media outlets as reported by Political Watch, while the y-axis represents the ideology of Twitter accounts. The dashed line represents the regression line.}
    \label{fig:app:twitter_media}
\end{figure}

\clearpage
\subsection{Clustering user voting behaviors}\label{app:sec:clustering}

To analyze user interactions based on their ideological positioning, we visualized two interaction matrices: one representing the total number of votes from users in ideological bin X to those in ideological bin Y, and another showing the average sign of those votes. The number of bins was determined by taking the square root of the total number of users (as there are $B×B$ interactions, where $B$ is the number of bins) and dividing by a normalization constant, which we set to 1.5.

To ensure comparability and reduce biases caused by variations in interaction activity, we applied an iterative normalization process to the total vote matrices. This process balances the influence of row and column totals, preventing highly active users from skewing the patterns. Specifically, each row and column of the matrix was normalized iteratively by dividing by its respective sum, ensuring each sum to one. This normalization reflects the relative strength of interactions independent of individual user activity levels, allowing for a fair comparison of voting behaviors and enabling meaningful clustering and visualization of user patterns across ideological groups.

We used the two matrices---normalized vote count and average vote sign---as inputs for a K-Means clustering algorithm. Each ideological bin was characterized by its normalized voting probabilities with all other bins. The optimal number of clusters was determined using the Variance Ratio Criterion (Calinski-Harabasz score) (\cite{calinski1974dendrite}), which evaluates the ratio of between-cluster dispersion to within-cluster dispersion.

The resulting clusters were directly used in visualizations. In cases where clusters overlapped (which occurred only at boundary bins), the overlap was removed to maintain clarity.

\clearpage
\subsection{Validation of political ideology} \label{app:validation}

To determine if the structural positioning of users in the \textit{user-to-user} network corresponds to left-right political ideology, we employed the following steps:

\subsubsection{Step 1: Binning User Positions}
Users were grouped into bins of 50 individuals each based on their structural positions as derived from the Signed Hamiltonian Eigenvector Embedding for Proximity (SHEEP) and Correspondence Analysis (CA) methods. This binning approach ensured that each bin represented a manageable and consistent sample size for subsequent analysis.

\subsubsection{Step 2: Calculating Voting Behavior}
For each bin, we analyzed the voting behavior of users by calculating how often users in that bin voted positively for each media outlet, compared to the average positive voting behavior across the platform. Let \( V_{u,m} \) represent the average voting sign of user \( u \) towards media outlet \( m \), and \( \bar{V}_m \) the average voting sign for \( m \) across all users. The deviation \( D_{u,m} \) was computed as:

\[
D_{u,m} = V_{u,m} - \bar{V}_m
\]

Positive deviations indicate a higher share of positive votes towards \( m \), while negative deviations indicate a lower share of positive votes.

\subsubsection{Step 3: Weighting by Media Ideology}
Each deviation  \( D_{u,m} \) was weighted by the ideological positioning of the media outlet \( I_m \), which was extracted from the \textit{user-to-news outlet} network (using Twitter ideology yields highly similar results). The weighted deviation for user \( u \) and outlet \( m \) is given by:

\[
W_{u,m} = D_{u,m} \cdot I_m
\]

This step provides an estimation of the ideological preference of user \( u \). Higher \( W_{u,m}  \) indicates a higher share of positive votes towards right-wing media or a lower share of positive votes towards left-wing media.

\subsubsection{Step 4: Adjusting for Voting Frequency}
To account for variations in the number of votes cast towards different outlets, the weight was further scaled by the ratio of the user's voting frequency \( F_{u,m} \) for outlet \( m \) to the average frequency \( \bar{F}_m \) across all users:

\[
R_{u,m} = W_{u,m} \cdot \frac{F_{u,m}}{\bar{F}_m}
\]

This adjustment ensures that outlets receiving disproportionately high or low attention are appropriately weighted in the analysis.

We calculate \( R_{bin,m} \) as the average \( R_{u,m} \) for all users in the bin and \( F_{bin,m} \) as the voting frequency of users in the bin toward outlet \( m \).

\subsubsection{Step 5: Aggregating Ideological Estimates at the bin level}
We aggregated the average weighted deviation for each bin, which was calculated as:

\[
I_{bin} = \frac{\sum_m R_{bin,m} \cdot F_{bin,m}}{\sum_m F_{bin,m}}
\]

This value represents the aggregated ideological positioning of users within the bin, based on their voting behavior towards media outlets.

\subsection{Voting behaviour and political ideology}\label{app:voting-behaviour-figures}
\begin{figure}
    \centering
    \includegraphics[width=\linewidth]{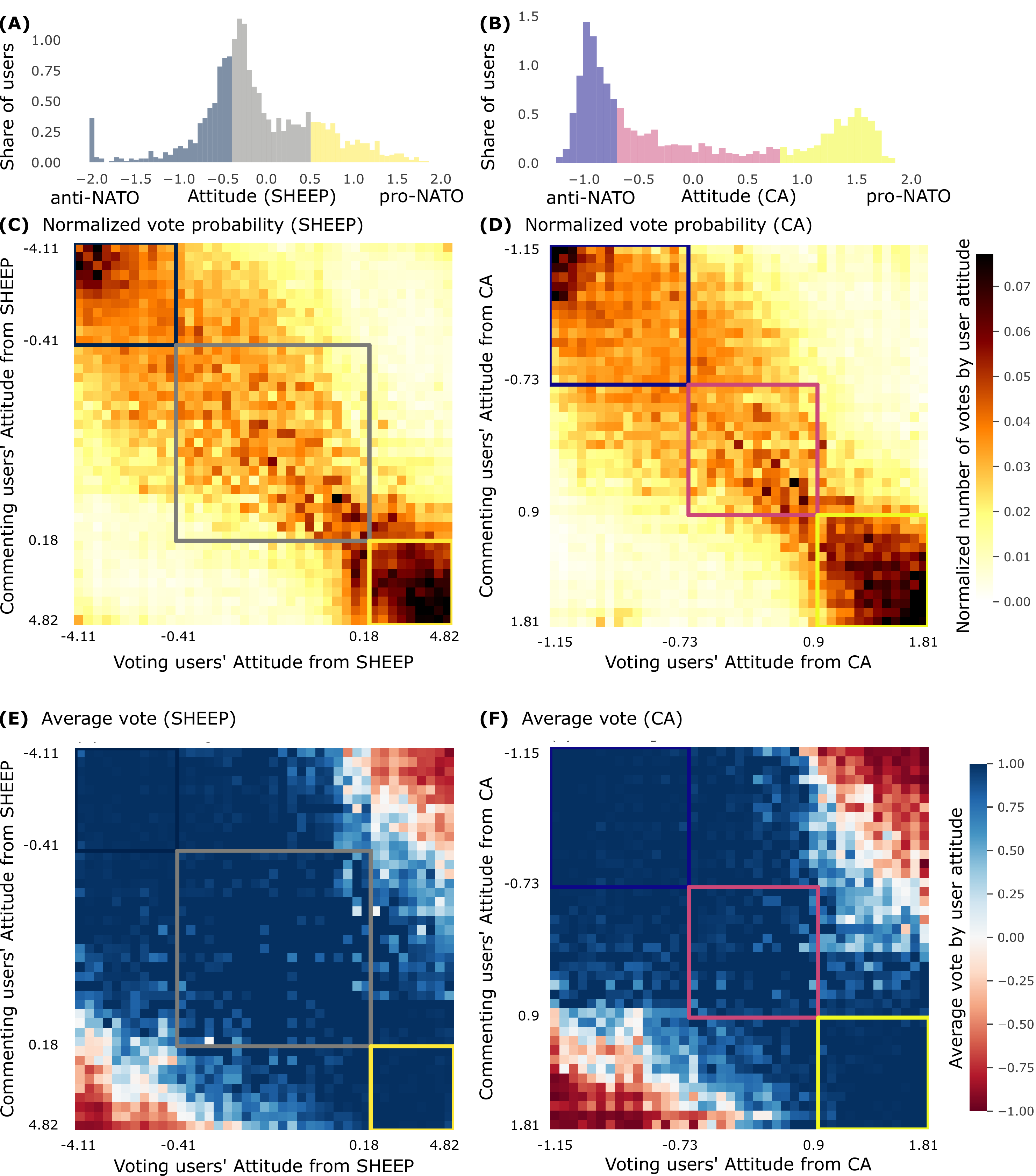}
    \caption{\textbf{Analyzing users' views on Russia-Ukraine war.} Panels A and B show the distribution of users across the embedding space for SHEEP and CA, respectively. Bar colors reflect k-means clustering (see~\ref{app:sec:clustering}). Note that histogram scales differ from Figure~\ref{fig:network_viz_russia} as the bins have uniform size. 
    Panels C and D display heatmaps of normalized vote probabilities, with rows representing voters' attitudes and columns representing the attitudes of users they vote on. Note that voting tends to occur between users with similar attitudes (votes often lie close to the diagonal).
    Panels E and F show average votes cast on stories and comments, ranging from -1 (all negative) to +1 (all positive). Users generally vote positively, except for extreme users, who downvote the opposite extreme.
    Clusters from k-means (matching panel A/B colors) are shown as boxes: anti-NATO (blue, dark purple), moderate (grey, pink), and pro-NATO (yellow). See~\ref{app:validation} for details on binning, vote normalization, and cluster interpretation.}
    \label{fig:users-russia}
\end{figure}

\begin{figure}
    \centering
    \includegraphics[width=\linewidth]{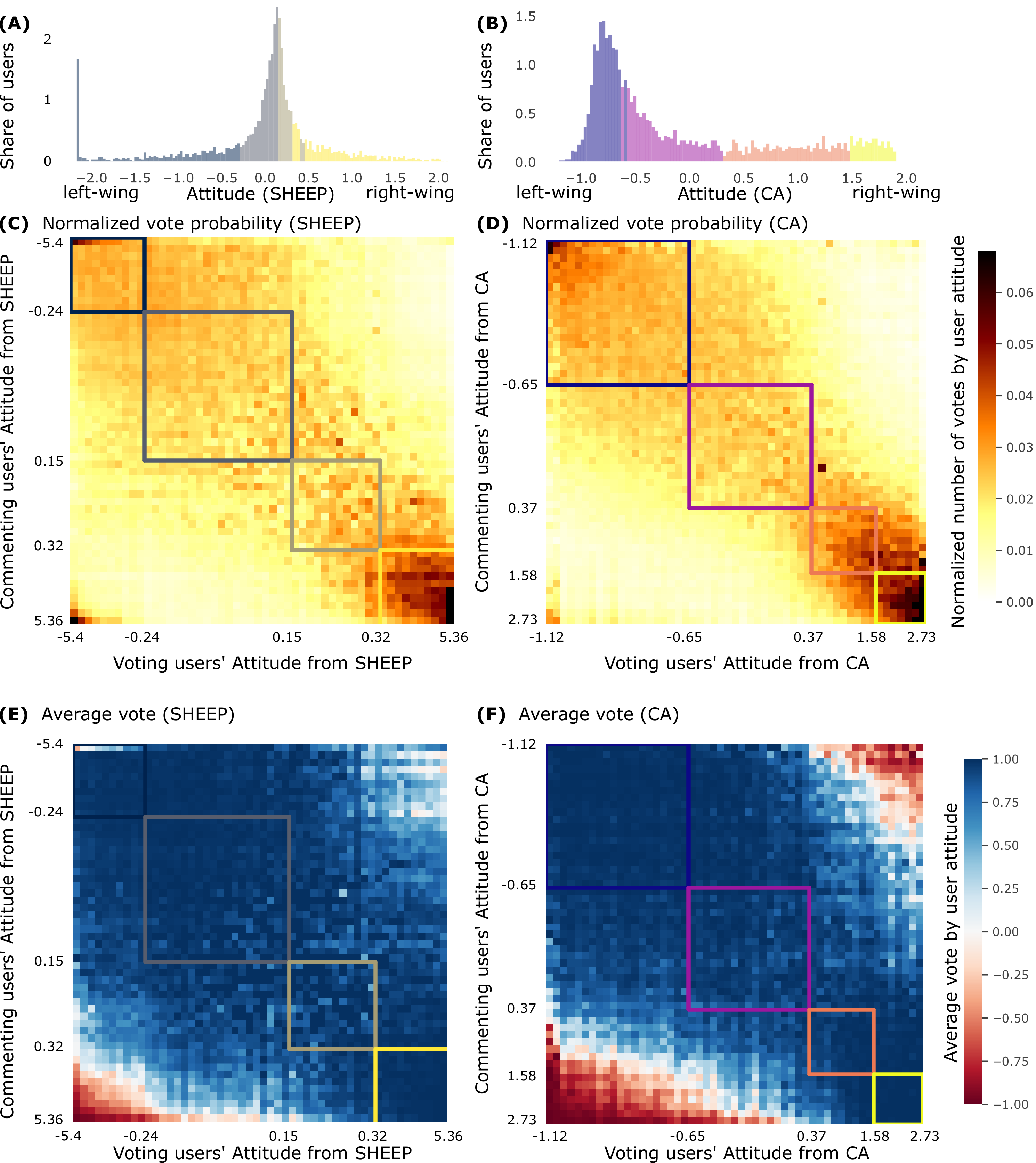}
    \caption{\textbf{Analyzing users' views on Broad Politics.} Panel A and B show again the distribution of users across the embedding space, for SHEEP and CA, respectively. Bar colors reflect k-means clustering(see~\ref{app:sec:clustering}). Note that histogram scales differ from Figure~\ref{fig:network_viz_politics} as the bins have uniform size. 
    Panels C and D display heatmaps of normalized vote probabilities, with rows representing voters' attitudes and columns representing the attitudes of users they vote on the users they vote on.
    Panels E and F show average votes cast on stories and comments, ranging from -1 (all negative) to +1 (all positive). Users generally vote positively, except for extreme left-wing users, who vote negatively towards the opposite extreme.
    Clusters from k-means (matching panel A/B colors) are shown as boxes: far-left (dark blue, dark purple), left-wing (light blue, pink), right-wing (grey, orange), and far-right (yellow). See~\ref{app:validation} for details on binning, vote normalization, and cluster interpretation.}
    \label{fig:users-politics}
\end{figure}

\clearpage
\subsection{SHEEP Null Model Comparison} \label{app:sheep-null}
SHEEP is a method designed to take signed networks as input, requiring both positive and negative edges to generate meaningful embeddings, while CA requires unsigned networks as input. In this section, we investigate whether the different embedding results we obtain from the two methods for the \textit{user-to-user} network are indeed due to the additional information gained from negative ties or just a result of the algorithmic choice. To that end, we create a signed network null model to input into the SHEEP algorithm, which obscures the information about the real negative ties by replacing all non-links in the unsigned network with an artificial negative link of weight -50, a similar magnitude to the largest positive links in the unsigned graph. This null model makes the same assumption as CA, that disconnected nodes are inherently dissimilar.  

\begin{figure}
    \centering
    \includegraphics[width=1\linewidth]{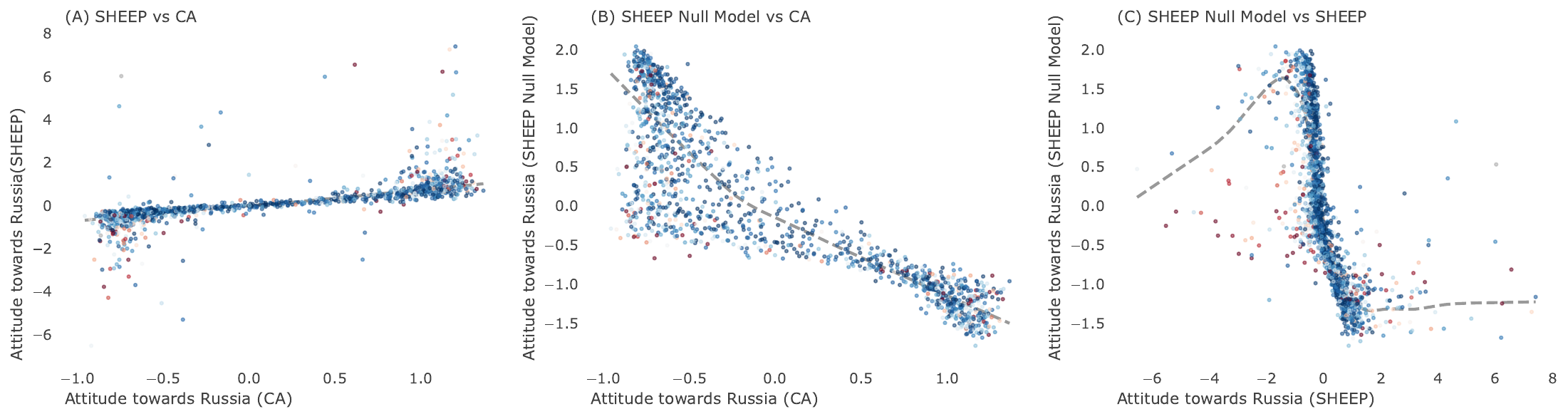}
    \caption{User-to-user network embeddings on Russia topic. Comparing user embeddings generated by CA, SHEEP, and SHEEP null model. As in Figure \ref{fig:network_viz_russia}, the color indicates the tendency of the user to vote positively (dark blue) or negatively (red). The Pearson correlation for SHEEP vs CA is 0.68, SHEEP null model vs CA is 0.83, and SHEEP null model vs SHEEP is 0.61. }    
    \label{fig:sheep-null-russia}
\end{figure}

\begin{figure}
    \centering
    \includegraphics[width=1\linewidth]{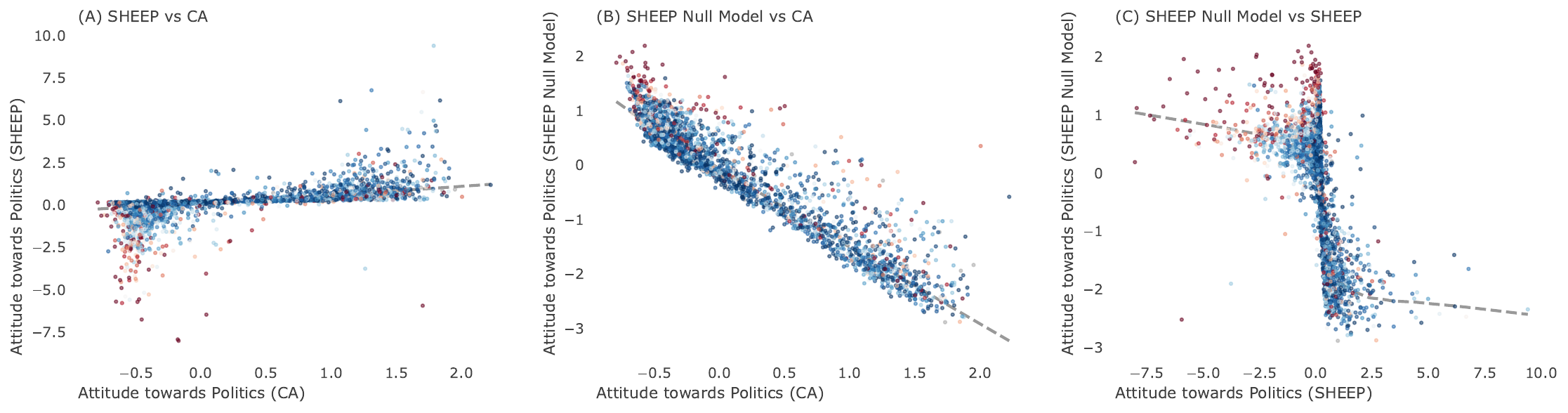}
    \caption{User-to-user network embeddings on Politics topic. Comparing user embeddings generated by CA, SHEEP, and SHEEP null model. As in Figure \ref{fig:network_viz_politics}, the color indicates the tendency of the user to vote positively (dark blue) or negatively (red). The Pearson correlation for SHEEP vs CA is 0.56, SHEEP null model vs CA is 0.94, and SHEEP null model vs SHEEP is 0.54. }    
    \label{fig:sheep-null-politics}
\end{figure}

Figure \ref{fig:sheep-null-russia} and Figure \ref{fig:sheep-null-politics} show the position of the user in the embedding space generated for the \textit{user-to-user} network across the three models. The color indicates the propensity of the user to vote positively (dark blue) or negatively (red) in the real signed network.  For both the Russia and Politics topic, we find that the embeddings generated by the SHEEP null model are extremely aligned with those of CA, much more than with the embeddings generated by SHEEP acting on the real signed network. In both cases, the Pearson correlation is highest between the SHEEP null model and CA (0.83 for Russia and 0.94 for Politics), indicating that the SHEEP null model produces an embedding that is highly linearly correlated with that produced by CA. These results once again support the conclusion that the differences we observe when incorporating negative ties are not simply artifacts of the algorithmic choice, but rather that the addition of the real negative tie information offers a more nuanced perspective on polarization. 

\end{document}